\documentclass[useAMS,usenatbib]{mn2e}
\usepackage[dvips]{graphicx}
\usepackage{subfigure,epsfig,longtable,multirow,amssymb,textcomp,multirow,capt-of}

\newcommand{\kms}{{km~s$^{-1}$}}
\newcommand{\teff}{{$T_{\mathrm{eff}}$}}

\newcommand{\msun}{M$_{\odot}$}

\title[The death of massive stars - II]{The death of massive stars - II. Observational constraints on the progenitors of type Ibc supernovae}
\author[J.J. Eldridge et al.]{John J. Eldridge$^{1}$ \thanks{E-mail: j.eldridge@auckland.ac.nz}, Morgan Fraser$^{2}$, Stephen J. Smartt$^{2}$,\and Justyn R. Maund$^{2}$, R. Mark Crockett$^{2}$\\
\\
$^{1}$The Department of Physics, The University of Auckland, Private Bag 92019, Auckland, New Zealand\\
$^{2}$Astrophysics Research Center, School of Mathematics and Physics, Queen's University Belfast, Belfast BT7 1NN, UK}

\pagerange{\pageref{firstpage}--\pageref{lastpage}} \pubyear{2012}
\begin{document}
\maketitle
\label{firstpage}

\begin{abstract}

The progenitors of many type II core-collapse supernovae have now been
identified directly on pre-discovery imaging. Here we present an
extensive search for the progenitors of type Ibc supernovae in all
available pre-discovery imaging since 1998. There are 12 type Ibc
supernovae with no detections of progenitors in either deep
ground-based or Hubble Space Telescope archival imaging. The deepest
absolute $BVR$ magnitude limits are between $-4^{m}$ and $-5^{m}$. We
compare these limits with the observed Wolf-Rayet population in the
Large Magellanic Cloud and estimate a 16 per cent probability we have
failed to detect such a progenitor by chance. Alternatively the
progenitors evolve significantly before core-collapse or we have
underestimated the extinction towards the progenitors.

Reviewing the relative rates and ejecta mass estimates from lightcurve
modelling of Ibc SNe, we find both incompatible with Wolf-Rayet stars
with initial masses $>25$\msun being the only progenitors. We present
binary evolution models that fit these observational
constraints. Stars in binaries with initial masses $\la20$\msun lose
their hydrogen envelopes in binary interactions to become low mass
helium stars. They retain a low mass hydrogen envelope until $\approx
10^4$ years before core-collapse; hence it is not surprising that
galactic analogues have been difficult to identify.

\end{abstract}


\section{Introduction}

Massive stars live fast and die young; core-collapse supernovae (SNe)
are the dramatic explosions which mark their deaths. During these
events a large amount of energy is deposited into the surrounding
interstellar medium, while elements heavier than boron are created
through explosive nucleosynthesis. The physical mechanism of
core-collapse in a star with an iron core is well established as the
main driving source behind the vast majority of core-collapse
explosions \citep {2012ARNPS..62..407J,2012arXiv1210.4921B}. However
the structure of the star before explosion and the variety in
explosions (kinetic energies, luminosities and variation in the
production of radioactive $^{56}$Ni) are still not well understood.
The observational characteristics are critically affected by the
evolutionary history of its progenitor star and uncertainties remain
(most notably in convection, mass-loss and rotation) that limit our
quantitative understanding \citep[see][ for a discussion of current
  limitations]{Lan12}.

Observed SNe are first divided into Types I and II, based respectively
on the absence or presence of hydrogen in the early-time optical
spectrum \citep{sntypes,snclassy}. These two broad types are further
divided into several subtypes. The most common Type II SNe are Type
IIP, characterised by a long plateau phase in their lightcurve of
constant luminosity, which lasts for several months. The lightcurves
of Type IIL SNe exhibit a linear decay, while Type IIb SNe only show
hydrogen in their spectra for the first few weeks before it
disappears. Type IIn SNe are slightly different, the hydrogen is seen
as narrow ($<$ 100 \kms) emission lines, which arise from interaction
with dense, optically thick circumstellar material. Type Ia SNe are
hydrogen and helium poor SNe that are believed to be thermonuclear
explosions arising as a result of accretion onto a carbon-oxygen white
dwarf and are not considered further here (see \citet{Hill00} for a
review). The other hydrogen-poor Type I SNe are subdivided into Types
Ib and Ic, which either do or do not show helium in their spectra
respectively. The type Ibc SNe\footnote{The types Ib and Ic are often
  difficult to distinguish and their classification is dependent on
  the phase of the spectra and signal-to-noise. Hence we will often
  use the common term of type Ibc to cover them jointly.}  have long
been associated with Wolf-Rayet progenitors
\citep[e.g.][]{Beg86,Gas86}.

In Paper I of this series, \citet{Sma09b}, the progenitors of Type IIP
SNe were identified as red supergiants (RSGs) with massive and
extended hydrogen envelopes. This conclusion was based on an analysis
of pre-explosion images (mostly from the {\it Hubble Space Telescope})
for twenty nearby SNe. These images were used to detect or place
limits on the luminosity and thus mass of the progenitors. For the
other subclasses of Type II supernovae there are some detections, but
insufficient numbers to consider in detail as an ensemble. There have
been three detected Type IIb progenitors for SNe 1993J, 2008ax and
2011dh, one transitional Type IIP/L progenitor for SN 2009kr and a
progenitor detected for the peculiar Type II SN 1987A
\citep{1987A,Ald94,2004Natur.427..129M,Cro08,2010ApJ...714L.254E,Fra10b,Mau11,2011dhARC}. Together
these observations suggest that if the progenitors of Type II SNe are
not RSGs, they are of a similar luminosity but with different surface
temperature ranging from blue to yellow supergiants.

Besides Type IIP SNe, the only other SN type for which there are a
large number of pre-explosion images available are the hydrogen
deficient Type Ibc SNe. There have been many attempts to directly
identify the progenitors of Ibc SNe in these pre-discovery images, but
no success see
\citet{Van03c,Mau05,2005ApJ...630L..29G,Mau05b,crockett07gr,Cro08}.  A
luminous outburst or eruption from a progenitor star for the type Ibn
SN2006jc was detected \citep{2007Natur.447..829P,2007ApJ...657L.105F},
but the progenitor was not detected in a quiescent phase.  The
progenitors of Type Ibc SNe could potentially be the classical,
massive Wolf-Rayet stars which are initially massive (M$_{\rm ZAMS}$
$\ga$25-30 \msun) but then lose their hydrogen envelopes through
radiatively driven stellar winds \citep{crowther2007}.  Such mass-loss
in massive stars can also be enhanced in massive binary systems.

Alternatively, it has been suggested that progenitors of Type Ibc SNe
could be helium stars with zero-age main-sequence (ZAMS) masses that
are too low to allow a {\em single} star to enter the WR stage. In
these cases, the star could have an initial mass of significantly less
than $\la25$ \msun. The lower initial mass limit for a SN progenitor
in such a system is uncertain, but is probably around the limit for
single stars to produce Fe-cores, or OMgNe cores, around
8\msun\ \citep{2009ARA&A..47...63S}. Such a star could lose its
hydrogen envelope through interaction with a binary companion, losing
mass by Roche-Lobe overflow or common-envelope evolution.
\citep[e.g.][]{1967AcA....17..355P,1985ApJS...58..661I,Pod92,1992ApJ...386..197T,1998A&A...333..557D,Nom95,2002ApJ...572..407B,Vanb2007}.
While theoretically such stars are likely progenitors of Type Ibc SNe,
there are no direct hydrogen-free analogues in our own galaxy, which
has led to some scepticism as to their existence. The closest such
systems we are aware of are V Sagittae systems, WR7a and HD45166,
which retain some hydrogen on their surfaces
\citep[e.g.][]{vsag,wr7a,hd45166a,hd45166b}.  However, recently
refined measurement of the relative rates of the SN types, together
with modelling of the evolution of binary systems, has added weight to
the argument that these binary stars represent a significant fraction
of Type Ibc progenitors
\citep[e.g.][]{1998A&A...333..557D,EIT,Yoo11,smithrates,ELT11}.

In this paper we combine the observed upper limits on the luminosity
of individual Type Ibc progenitors with theoretical modelling of a
binary population to constrain the progenitor population as a
whole. In particular, we aim to determine whether the observed
magnitude limits for Type Ibc SN progenitors (and their relative
rates) can be reconciled with the observed population of WR stars in
the Milky Way and Magellanic Clouds, or whether they are better
matched by a model population including binary and single stars.

\section{The local SN rate}
 \label{s2}
 
We have followed the methodology of Paper I in sample selection. The
list of supernovae maintained by the IAU Central Bureau for
Astronomical
Telegrams\footnote{http://www.cbat.eps.harvard.edu/lists/Supernovae.html}
was searched for all supernovae with a named host galaxy discovered in
the fourteen year period from 1 Jan 1998 to 30 March 2012.  This
sample was then cross matched against the the HyperLEDA galaxy
database\footnote{http://leda.univ-lyon1.fr/} to identify all
supernovae for which the host galaxy had a recessional velocity
(corrected for Local Group infall on Virgo) of 2000~\kms\ or less,
corresponding to a distance limit of 28 Mpc for H$_0$=72 \kms
Mpc$^{-1}$. For SNe where there was not a named host in the IAU
catalogue, HyperLEDA was searched for any galaxy within 1.5\arcmin of
the SN position. Finally, the coordinates of any SNe which did not
have a named host or a galaxy within 1.5\arcmin, and was brighter than
magnitude 16 were queried via the NASA Extragalactic Database
(NED)\footnote{http://ned.ipac.caltech.edu/}. We also cross-checked
our sample against the catalogue of \citet{Len12}, by searching for
all SNe within 35 Mpc, and manually checking LEDA for the distance to
any which were not in our sample. We did not identify any SNe within
our distance limit which we had missed, although we note that several
SNe (e.g.. SNe 2001fu, 2004ea and 2006mq) which were within our
distance limit do not have an associated distance in the Lennarz et
al. catalogue.

From this, we obtained a sample of 203 transients over the 14.25 year
time period within a 28 Mpc volume, as listed in Appendix
\ref{a1}. For comparison, in Paper I we found 141 transients within
the same volume over the 10.5 years from 1 Jan 1998 to 30 June 2008,
which appears broadly consistent with the number found here : 13.4 SNe
yr$^{-1}$ in Paper I, compared to 14.3 SNe yr$^{-1}$ over the extended
14.25 yr period that we consider here. Most of the SNe in the sample
have a spectral typing, which are listed in Appendix \ref{a1} together
with the recessional velocity of the host. Those SNe which are of
uncertain type, or which are otherwise noteworthy, and which were not
discussed in \citet{Sma09b}, are discussed individually below.

\begin{enumerate}

\item{SN 2010dn

The nature of this transient is still
debated. \citet{2010CBET.2300....1V} claimed it is an eruptive
outburst from a dust-enshrouded LBV, but the spectra bear resemblance
to the SN2008S-like transients that are still potentially SNe.
\citep{2009MNRAS.398.1041B,Smi09,Pri10}. \citet{Ber10} set upper
limits on the progenitor magnitude of SN 2010dn from archival Spitzer
and Hubble Space Telescope (HST) observations of the site of the
transient. \citet{Smi11} presented a re-analysis of the same HST data,
and similarly to Vinko et al. favoured a supernova impostor / LBV
eruption as the most likely explanation.  \citet{Hof11} report that a
MIR detection of 2010dn which is again similar to that found for other
SN2008S-like events. We thus leave it in the bin of ``Uncertain or
non-supernova'' in Table\,A1.  }

\item{SN 2009ip

 As noted in Table \ref{table:fullsample}, the original discovery of
 the object we call SN2009ip was an LBV outburst. \citet{Smi10} and
 \cite{Fol11} identified the progenitor as a star of $\sim$60
 \msun\ using archival HST images. A similar transient, UGC 2773
 OT2009-1 was discussed by Smith et al., and again was found to have a
 massive LBV progenitor.  These LBV outbursts or eruptions are not
 included in our rate calculations but are listed, insofar as we are
 aware of their existence, in Table \ref{table:fullsample}).  The star
 which underwent an LBV eruption to give SN2009ip may now have
 produced a supernova event, during August-September 2012.  If
 confirmed this is perhaps one of the most interesting supernovae ever
 to have occurred and is already the focus of intensive study
 \citep{2012arXiv1209.6320M, 2012arXiv1210.3568P, 2012arXiv1210.3347P,
   2012arXiv1211.4577L,2013MNRAS.tmp.1499F,Mar13}. The event occurred outside our timeframe,
 hence we do not include it as a SN IIn for rate estimates. However,
 we recognise that SN 2009ip has an extremely important role in
 linking the progenitor star, SN explosions and the explosion
 mechanism.}

\item{SN 2008iz

The SN was discovered at radio wavelengths \citep{Bru09},
 and has been the subject of extensive radio follow-up
 \citep{Bru10,Mar10}. \citet{2009ATel.2131....1F} observed the SN with
 Gemini+NIRI, and initially claimed that the SN could not be seen in
 the NIR. However a careful reanalysis by \cite{2012arXiv1207.1889M} 
 shows that the SN can be seen when image subtraction techniques are
 used.  They find a likely very high line of sight extinction, but
 probably  less than $A_{V}\approx 10$. As this was a SN which was not, 
and could not, have been discovered by current optical surveys we 
do not consider it for our rate estimates. Its type is also currently
unknown which would not enhance our measurements. It does 
serve as a reminder on the number of SNe that may be occurring in 
high star-formation rate locations which are extinguished
\citep{2012arXiv1207.1889M,2012ApJ...756..111M,2012A&A...537A.132B}. 
}

\item{SNe 2008ge, 2008ha, 2010ae \& 2010el.

 The SNe have all been identified as SN 2002cx-like. SN 2008ha is
 perhaps the most unusual SN considered in the sample; it showed low
 velocities and a faint absolute magnitude, and its interpretation is
 still contentious. \citet{2009Natur.459..674V} suggested that the SN
 was the faint core-collapse of a massive progenitor on the basis of
 its spectral similarity to SN 2005cs, however
 \citet{2009AJ....138..376F} and \citet{Fol10a} have questioned this
 interpretation and proposed that it was in fact a thermonuclear
 SN. Foley et al. argue that intermediate mass elements, whose
 presence Valenti et al. claimed was counterfactual to a Type Ia
 scenario, can be present in a thermonuclear supernova. Valenti et
 al. noted that SN 2008ha appears to be related to the class of SN
 2002cx-like SNe \citep{Li03}, and suggested that all of these may be
 related to Type Ibc SNe, possibly with fallback of the inner most
 ejecta onto the compact stellar remnant. While a fallback explosion
 model has been constructed for SN 2008ha \citep{Mor10}, it requires a
 progenitor which has a mass lower than that typically thought to
 undergo this process \citep{Heg03}. Remarkably, three other SNe
 besides SN 2008ha in this sample have been suggested to be SN
 2002cx-like. The discovery of this many SN 2002cx-like SNe in such a
 short period could suggest that a significant number of these events
 have gone undetected previously, likely due to their faint
 magnitude. For one of them, SN 2008ge, \citet{2010AJ....140.1321F}
 have analysed pre-explosion data, and find no evidence of either
 ongoing star formation, or the young massive stellar population which
 would be expected to be found in the environment of a core-collapse
 SN. While the nature of these SNe is still open to debate, at present
 the balance of evidence seems to favour a thermonuclear
 interpretation, and so we have excluded them from the core-collapse
 sample on this basis. However, given recent evidence that the class
 of SN 2002cx-like SNe may be significantly more heterogeneous than
 previously thought \citep{Nar11} and may not all share the same
 explosion mechanism, a more detailed search for the progenitors of
 these SNe would be a worthwhile endeavour. }

\item{ SN 2008eh

As far as we are aware, there is no published spectrum of SN
2008eh. Hence we have followed the lead of \citet{Hor11} and assumed
the SN is a core-collapse event (probably Type Ibc) on the basis of
its absolute magnitude and lightcurve, and proximity to a H{\sc ii}
region in a spiral galaxy. Nonetheless, we have excluded this SN from
the rate calculations.}

\end{enumerate}

\begin{figure}
\begin{center}
\includegraphics[width=0.8\columnwidth,angle=0]{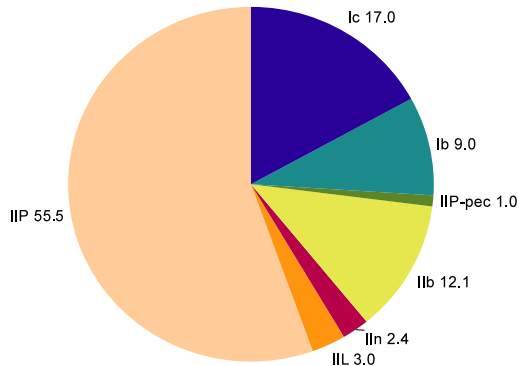}
\caption{Percentage of CCSNe which are of a particular type as found in this work.}
\label{fig_rates}
\end{center}
\end {figure}

We summarise the relative rates of different core-collapse SNe within
our distance limit in Table \ref{table:fullsample}. Discarding all SNe
which were designated as Type Ia (55 SNe, or 30.2 per cent of the
total classified sample) along with SN impostors and unclassified
transients, we find relative rates for the various types of
core-collapse SNe as listed in Table \ref{table:rates} and shown in
Figure \ref{fig_rates}. SNe which were classified as a ``Type Ibc''
were divided between the Type Ib and Ic bins, according to the
observed ratio of these subtypes (9:17). Type II SNe were distributed
between the IIP, IIL, IIb and IIp-pec bins in the ratio
55:3:12:1. Uncertainties were calculated from the standard error on a
Poissionian distribution (i.e. as $\sqrt{N}$). We have made no attempt
to further subdivide the Type IIb SNe into the ``compact'' and
``extended'' subtypes proposed by \citet{Che10}. Although there is
some scepticism as to whether these subtypes exist
\citep{2013ApJ...762...14M}.

\begin{table*}
\caption[]{The relative frequency of core-collapse SN types discovered
  between 1998-2012.25 (14.25 yrs) in galaxies with recessional
  velocities less than 2000 kms$^{-1}$, as listed in
  Table\,\ref{table:fullsample}. The values in parentheses in the
  Number column include SNe of indeterminate subtype as discussed
  in the text.}
\label{table:rates}
\begin{tabular}{rccc}
\hline
\hline         
SN Type 		& Number		& Relative	 rate
(per cent)	& LOSS (per cent) \\
\hline
IIP     	          	& 55  (70.5) 	& 55.5 	$\pm$ 6.6 &  48.2 $^{+5.7}_{-5.6}$\\		
IIL     			& 3    (3.8)	& 3.0	$\pm$ 1.5 &  6.4 $^{+2.9}_{-2.5}$\\ 
IIn      		        & 3    (3)	& 2.4	$\pm$ 1.4 &  8.8 $^{+3.3}_{-2.9}$\\
IIb      		        & 12  (15.4)	& 12.1 	$\pm$ 3.0 &  10.6 $^{+3.6}_{-3.1}$\\
IIpec (87A-like)	        & 1    (1.3)	& 1.0 	$\pm$ 0.9 &  ... \\
Ib      			& 9    (11.4)	& 9.0 	$\pm$ 2.7 &  8.4 $^{+3.1}_{-2.6}$\\
Ic      			& 17  (21.6)	& 17.0 	$\pm$ 3.7 & 17.6 $^{+4.2}_{-3.8}$ \\
\hline 
Total			& 100 (127)	& 		\\
\hline
\hline
\end{tabular}
\end{table*}

Our relative rates compare favourably with those found in Paper I, the
most noticeable difference being the increase in Type IIb SNe at the
expense of Type Ibc events. A possible explanation for this is that
SNe are now discovered (on average) sooner after explosion, and so
Type IIb SNe which previously would have been detected late and
classified as a Type Ibc are now correctly identified.

\citet{smithrates} measured SN rates from the Lick Observatory
Supernova Search (LOSS) sample, based on those SNe selected for the
luminosity function analysis by \cite{2011MNRAS.412.1441L}, with a
total of 106 CCSNe within 60\,Mpc.  This is smaller than we present
here, but is a much more homogeneous sample and is a well controlled
SN survey. We emphasise that we are assimilating all discoveries and
not carrying out a controlled SN search, but the results should be
compared given the large numbers available.

Smith et al. did not separate out the IIpec category, but did separate
the Ibc-pec SNe. To facilitate a comparison between our rates, we
re-distributed the Type Ibc-pec SNe from Smith et al. between the Ib
and Ic categories (weighted by the Type Ib:Ic ratio), and the Type
IIpec SNe from this work between the Type IIP, IIL and IIb
categories. We find a good agreement between the Smith rates and those
presented in this paper for the Type IIb, Ib and Ic SNe. We find a
Type IIP rate which is 7.3 per cent higher than that derived by Smith
et al, although the uncertainties of Smith et al. and this work do
overlap. If we combine the Type IIL and IIP rates, then the
discrepancy is smaller, raising the possibility that some of our Type
IIP SNe may in fact be Type IILs (and vice-versa for Smith et
al.). Smith et al. suggest that the LOSS rates are more reliable due
to the SNe all coming from the same survey with well controlled and
known selection effects and good spectroscopic and photometric
coverage.  The latter is certainly important to distinguish between
II-P and II-L SNe. In Paper\,I we attempted to classify II-P SNe from
literature, archive and web data searches and classified any type II
SN that had a constant magnitude phase for more than 30\,days a type
II-P. This of course is not as homogeneous a dataset as that from
LOSS.  There are 23 SNe in common between the Smith et al. sample and
that presented in Paper\,I (there are no extra SNe to add from
2008-2012, as the LOSS sample stops in 2006).  Of those, 19 are
identical classifications and two have type II classifications by us
and II-P classifications by LOSS.  Including those two updated LOSS
classifications here would make the II-P discrepancy worse.  The other
two were fairly trivial differences : Ic vs Ibc/IIb (2004ao) and Ibc
vs Ib (2004qn) in Paper\,I and LOSS respectively. Hence there is no
evidence we can see that points to systematic miss-classifications in
the smaller volume sample presented in Paper\,I and supplemented here.

The only SN type for which the rates and their uncertainties do not
overlap at all are the Type IIn SNe, where Smith et al. find a higher
rate than we do (8.8$^{+3.3}_{-2.9}$ per cent versus 2.4 $\pm$ 1.4 per
cent). Smith et al. attribute this to the fact that LOSS has better
and more extensive spectroscopic and photometric coverage. This is
difficult to assess as the global SN community certainly has extensive
resources to bear on SNe within 28\,Mpc. Classification results are
rapidly reported, data are published on reasonable timescales, and we
have made attempts to search archives and request private information
where we can; hence it's not clear to us that this statement in Smith et
al. is necessarily correct. The Li et al. sample was produced for
luminosity function work, which requires SNe to be ``in season'' and
full lightcurve coverage from before explosion. This is essential for
luminosity function measurements, but not for rate estimates.  For
example, long duration plateau SNe may be preferentially excluded as
their ``season'' lasts longer than other types. It's still possible
that the high IIn rate in LOSS is due to them being brighter and being
preferentially discovered between 30-60\,Mpc compared with $<28$\,Mpc.
However overall, the agreement is reasonable despite the differences
in the construction of the two samples.

As described in Paper\,I for all SNe, we searched the HST archives
using the Querator\footnote{http://archive.eso.org/querator/} and
MAST\footnote{http://archive.stsci.edu/} webpages.  For core-collapse
SNe we find that out of 127 SNe, 41 of them had HST images taken
before discovery (i.e. 32 per cent) with the SN falling on the HST
camera field of view.  For type Ia SNe the fraction was 20 out of 55
(i.e. 36 per cent).  Many more host galaxies of SNe had observations
taken before explosion, but the SNe fell outside the small fields of
view of the HST cameras (16 per cent of core-collapse and 16 per cent
of Type Ia ; see Table \ref{table:fullsample}). We also searched the
archives for the ESO telescopes\footnote{http://archive.eso.org/cms/},
Gemini North and South
telescopes\footnote{http://cadcwww.dao.nrc.ca/gsa/}, the Subaru
Telescope\footnote{http://smoka.nao.ac.jp/}, the William Herschel
Telescope
(WHT)\footnote{http://casu.ast.cam.ac.uk/casuadc/archives/ingarch} and
the Canada-France-Hawaii Telescope
(CFHT)\footnote{http://cadcwww.dao.nrc.ca/cfht/} for any SNe in a host
with a recessional velocity of $<$1000 \kms. There are other 8 meter
class telescopes which could be of use for progenitor searches, such
as the Keck telescopes and the Large Binocular Telescope, however the
absence of publicly accessible data archives precludes this.

\section{Estimating progenitor masses from SN rates}
\label{s3}

\citet{smithrates} used their observed SN rates to estimate progenitor
mass ranges, given a standard IMF and assuming that more massive stars
will undergo greater mass loss. They reached the conclusion that a
simple single star initial mass prescription is unable to reproduce
the observed SN rates, and that binary evolution is a crucial factor
for Type Ibc SNe. This is based on the relatively high rates of Ibc
SNe and the fact that the mass range for classical, massive WR stars
cannot reproduce the rates from Salpeter IMF arguments \cite[also see
discussion in][]{2009ARA&A..47...63S}. This problem was originally noted 
by \cite{Pod92} and \cite{1995PhR...256..173N} who proposed that
around 30 per cent of stars above
8\msun\ could lose mass in interacting binaries and produce the Ibc 
SN population.  Here we use a more detailed stellar population model 
to reproduce the observed rates. The main difficulty is attaching an 
observed SN type to the end point of the stellar model and the physical
mass of H and He left in the model stars. 

To reproduce the SN rates with a binary population, we use synthetic
stellar populations from the Binary Population and Spectral Synthesis
code (\textsc{BPASS}). This code is able to create stellar populations
of both single and binary stars, and is described in detail in
\citet{EIT}, \citet{ES09} and \citet{ELT11}. In these papers it is
shown that populations including binaries are better able to reproduce
observed stellar populations, both resolved and unresolved.

We have used the BPASS code to predict the relative rates of different
types of SNe determined in Section \ref{s2} and listed in Table
\ref{table:rates}. Our fiducial progenitor populations are a solely
single-star population, and a mixed binary/single-star population with
one single star for every binary. The models are computed for two
metallicities, $Z=0.008$ and 0.020, corresponding to the Large
Magallenic Cloud abundance and the Solar abundance respectively. We
average over these values as the SNe in our sample occur in nearby
spiral galaxies which will have metallicities between these two
extremes (see discussion in Paper\,I).  We also note that at lower
metallicities quasi-homogeneous evolution may become important
\citep{yoon,ELT11} but we do not currently include this pathway. We
consider all stars from $5$M$_{\odot}$ to $120$M$_{\odot}$ in both
single and binary populations. For the binary populations we include a
range of mass ratios for the secondary star ($q=0.3$, 0.5, 0.7 and
0.9) and a range of initial separations from $\log (a/R_{\odot})=1$ to
4. According to our models, approximately two-thirds of these binaries
interact. This is consistent with the recent observations of
\cite{2012Sci...337..444S}, although those are measurements of O-stars
with ZAMS masses of 20\msun\ and greater.

The precise configuration of a star at the moment of core-collapse,
and how this determines the resulting SN type is still quite uncertain
\citep[e.g.][]{Heg03,ETsne}. Stellar evolution codes calculate model
stars with quantitative residual envelope masses of H and He and are
given SN types by assuming values of the minimum masses required to
give SNe of types II and Ib respectively \cite[for example, see the
  discussion in][]{dessart}; hence we look at this problem in a different
way. We insist that the model populations must reproduce the observed
relative SN rates in Table \ref{table:rates}, as in
\citet{ELT11}. Each star in the ensemble is evolved to the end of core
carbon burning.  At this point, a resulting SN type is assigned to the
model, based on the mass of the hydrogen envelope, the H to He mass
ratio, and the fraction of He in the ejecta. Then we can determine
what the minimum mass of H is to produce the total numbers of type II
and type II-P populations respectively.  As suggested by
\cite{smithrates}, we find that a single star population cannot
produce the observed rates of Ibc SNe

A mixed population of binary and single stars (with one single star
for every binary) produces end stellar points that can match the
relative rates of type Ibc vs. II SNe {\em if} we assume that the
final hydrogen and helium masses given in Table \ref{snpram} are what
is required to produce the various types.  We find that a Type II SN
requires at least 0.003\msun\ of H in the ejecta, while a Type IIP SN
must have a H to He mass ratio of greater than 1.05 and that the total
mass of hydrogen is greater than 1.5~\msun .  While this mass of
0.003\msun\ of H is at first sight low, it would be the minimum for a
IIb SNe.  The synthetic spectra calculations of \cite{dessart} show
that even 0.001\msun\ can produce visible hydrogen lines (H$\alpha$ in
particular), as long as the surface mass fraction is higher than 0.01.

For a Type Ib SN we require the mass fraction of helium in the ejecta
to be greater than 0.61, while for a Type Ic the He fraction must be
less than this. {In our Ibc progenitor models we find the
  ejecta is dominated by either helium or carbon and oxygen. Therefore
  for a large ejecta mass a large mass of helium would be required for
  a Ib. We find few stars with such ejecta masses in the models and
  they all typically have a few times $0.1M_{\odot}$ of
  helium. Considering the typical ejecta masses for Ibc SNe of
  $2--3M_{\odot}$ our required ratio seems appropriate. We note that
  it is likely that helium mass alone does not determine if a SN is
  type Ib or Ic. Other factors such as the mixing of the ejecta are
  important as discussed by \citet{2012MNRAS.424.2139D}.}

Comparing these mass values to those in \citet{ELT11} - which were
based on the rates of Paper I - we find that they have changed by a
similar percentage as the SN rates between Paper I and this paper. The
most uncertain value is the amount of hydrogen that can be hidden in a
type Ibc SNe, the value here being lower by a factor of 17. This
difference is chiefly due to the larger fraction of Type IIb SNe in
this work, which causes the minimum required H mass for a Type II SN
to be lower.

We appreciate that this fairly simple and well defined classifications
of the  progenitor models are, in reality, complicated by two
factors.  \cite{2002ApJ...566.1005B}  show that some Ib SNe may well contain some
residual hydrogen, given good quality spectra and model photosphere
fits. \cite{2012MNRAS.424.2139D} discuss that  it may be quite possible to get a 
Ic SN with helium in the progenitor star, when He\,{\sc i} is not
excited due to weak mixing of $^{56}$Ni into the He rich regions. 

In summary we can reproduce the observed rates in Section 2, with a
mixed population of single stars and binaries with one single star for every
binary system, as long as we attach the resulting progenitor stars into 
SN classification bins with the envelope mass fractions as reported in 
Table \ref{snpram}.  This means that one third of all progenitors have
had their envelopes stripped by binary interactions via either Roche
lobe overflow or common envelope evolution \citep[consistent with
results found in][]{smithrates}. 

\begin{table}
\caption[]{The required parameters for a star to give rise to a specific SN type assuming all SN are observable.}
\label{snpram}
\begin{tabular}{lcccccc}
\hline
\hline
SN 	 	& Final              			&$M_{\rm CO core}$ 	&  $M({\rm H})$ 	&  $M({\rm H})$  	&  $M({\rm He})$  	\\
Type  	& Mass$/M_{\odot}$    	& $  /M_{\odot}$  		&  $/M({\rm He})$ 	&$  /M_{\odot}$    	&$/M_{\rm ejecta}$   \\
\hline
IIP          	&$> 2$				&$> 1.38$				&$\ge 1.05$		& $>1.5$		& -				\\
II           	&$> 2$				&$> 1.38$				&$< 1.05 $  		&  $>0.003$		&-  				\\
Ib           	&$> 2$				&$> 1.38$				&--    			& $\le 0.003$   	& $\ge 0.61$		\\
Ic           	&$> 2$				&$> 1.38$				&--   			& $\le 0.003$   	&  $< 0.61$ 		\\
\hline
\hline
\end{tabular}
\end{table}

Finally it has been suggested that SNe that form black holes directly
do not produce any SN display. \citet{Heg03} estimate that there will
be no optical display if the final helium core mass is greater than
15M$_{\odot}$; in our models this corresponds to a remnant mass of
8M$_{\odot}$, although this idea has be called into question by recent
results from \citet{2012ApJ...757...69U}. We have created a synthetic
population in which we remove all core-collapse events with a remnant
mass $>8$\msun\ (approximately 10 per cent of all SNe). Even without
these events, we find that the same parameters in Table \ref{snpram}
can be used to reproduce the observed SN rates, so long as we increase
the fraction of binaries in the population.

\section{Limits on Type Ibc SN progenitors from archival imaging}

Alongside estimates from the relative rates of CCSN types, direct
detections and upper limits on progenitors from archival data can help
constrain their progenitor population. In the following section we
summarise the published limits for Type Ibc progenitors within a
recessional velocity limit of 2000 \kms, and extend this sample with
limits on an additional six SNe.

The general technique used is the same as that described
in \citet{Sma09b}. We align a pre-explosion image to a post-explosion
image by means of background stars in the vicinity of the SN. We then
search for a source coincident with the SN in the pre-explosion
image. If a coincident source is identified, then its magnitude can be
measured, otherwise (as in all of the following cases) either pixel
statistics or artificial star tests are used to set a limiting
magnitude for a non-detection. All magnitudes quoted here are
Vegamags.

\begin{table}
\caption{HST Observational Data}
\label{tab_obsdata}
\begin{tabular}{lccc}
\hline
\hline
Filter    & Date  &    Exposure time &     \\
\hline
SN2001ci  &                                     &                      &          \\
WFPC2 & F547M  	& 1999-03-04 	 	& 320		\\
WFPC2 & F606W	& 2001-01-21  		& 560 		\\
WFPC2 & F658N	& 1999-03-04 		& 1200		\\
WFPC2 & F658N	& 1998-11-26 		& 8900		\\
WFPC2 & F814W  	& 1998-11-26 	 	& 800		\\
WFPC2 & F814W  	& 1999-03-04  		& 140		\\
\hline
SN2003jg  &                                     &                      &          \\
WFPC2 & F450W	& 2001-08-02  		& 460 	 \\
WFPC2 & F606W	& 1994-06-21 		& 160	 \\
WFPC2 & F814W  	& 2001-08-02 	 	& 460	 \\
\hline
SN2004gn  \\
WFPC2 &F658N	& 2011-03-02		& 1400   	 \\
WFPC2 &F606W	& 				& 600 	 		 \\
\hline
SN2005V  \\
ACS  	&  	F658N	& 2004-04-10  		& 700 		 		 \\
	  	& 	F814W	& 				& 120 		 		 \\
NICMOS	&	F160W	& 2002-12-04		& 96					\\
		&	F187N	&				& 640				\\
		&	F190N	&				& 768				\\
WFPC2	&	F450W	& 2001-07-06		& 460				\\
		& 	F814W	&				& 460				\\
		& PC	F606W	& 2001-01-05		& 560				\\	
\hline
\end{tabular}
\end{table}

\subsection{2000ew}

SN 2000ew was discovered by amateur astronomers \citep{Puc00}. The SN
was initially spectroscopically classified as a Type Ia SN
\citep{Den00}, before \citet{Fil00} obtained a second spectrum which
showed the SN to be of Type Ic. A Tully-Fisher distance of 18.2 Mpc
has been measured for the host galaxy of SN 2000ew, NGC 3810. As the
coordinates of NGC 3810 are on the outskirts of the Virgo cluster
(11h41m, +12$^{\circ}$28\arcmin) the kinematic distance estimate (13.4
Mpc) may not be reliable. We hence adopt the Tully-Fisher distance as
a more conservative estimate.

\citet{Mau05b} set a limit of F606W$>$24.6 on the progenitor of SN
2000ew from pre-explosion WFPC2 imaging, while \citet{Van03} find a
similar limit of F606W$>$24.7. Maund \& Smartt estimated the reddening
towards the SN from nearby stars to be E(B-V)=0.01, but as this is
less than the foreground value from NED (E(B-V)=0.044) we adopt the
latter as did Maund \& Smartt. The F606W absolute magnitude is listed
in Table\,\ref{obsprogenitors}.

Late time NIR spectra of SN 2000ew were presented by
\citet{2002PASJ...54..905G}, these can help further elucidate the
nature of the progenitor. Carbon monoxide (CO) emission was observed
in the spectrum, albeit at a lower velocity ($\sim$2000 \kms) than
might be expected for a progenitor with a 2.1 \msun C+O core
\citep{Iwa94}. While the CO velocity could be interpreted as arising
from a lower energy explosion, perhaps pointing to a less massive
progenitor, it may also be explained with an asymmetric explosion with
a largest component of the velocity perpendicular to the line of
sight. We also note that \citet{2002PASJ...54..905G} also found narrow
He{\sc i} and [Fe{\sc ii}] lines, but no Brackett series lines, in
their spectrum which they attribute to H-poor material shed by the
progenitor star prior to core-collapse.

\subsection{2001B}

SN 2001B was discovered in IC 391 by \citet{Xu01}, and initially
classified as a Type Ia SN by \citet{2001IAUC.7563....2M} before
\citet{2001IAUC.7577....2C} reclassified the SN as a probable Type
Ib. A lightcurve was presented by \citet{2006PZ.....26....3T} which
reached a maximum of -17.5 in $V$, supporting the classification of SN
2001B as a Type Ib. IC 391 has a Tully-Fischer distance of 25.5 Mpc,
and a recessional velocity distance of 25.5 Mpc (from NED).

\citet{Van03c} claimed the tentative detection of a progenitor
candidate for SN 2001B from an alignment to a ground-based natural
seeing image. Subsequently \citet{Mau05b} presented late time HST
imaging showing that the SN was offset from the candidate. Maund \&
Smartt set an upper limit of F555W=24.30$\pm$0.15 for the progenitor
of SN 2001B, which is in agreement with the limit suggested by Van Dyk
et al.  if their progenitor candidate was {\it not} the
progenitor. Maund \& Smartt estimate a reddening of
E(B-V)=0.102$\pm$0.030 from nearby supergiant stars, which is slightly
lower than the line of sight extinction of E(B-V)=0.14 from NED, we
hence adopt the latter as the larger value. The absolute magnitudes are
reported in Table\,\ref{obsprogenitors}.

\subsection{2001ci}

SN 2001ci was discovered as part of the Lick Observatory and Tenagra
Observatory Supernova Searches (LOTOSS) by \citet{Swi01}. It was noted
at discovery that the SN had a low absolute magnitude, and it was
originally suggested that the transient may be a SN
impostor. \citet{2001IAUC.7638....1F} subsequently obtained a spectrum
of the SN, and classified it as a highly reddened Type Ic SN, with a
line-of-sight extinction of $A_V\approx5$--6 mag.

While NGC 3079 had been observed with HST prior to the SN explosion as
detailed in Table \ref{tab_obsdata}, the images are not sufficiently deep
to have a reasonable expectation of seeing a progenitor given the high
extinction. In addition, we were unable to locate any high resolution
post-explosion imaging for SN 2001ci. NGC 3079 was observed with
HST+WFPC2 in the F300W filter on 2001 Dec 9 ($\approx$6 months after the
SN explosion) and the SN coordinates fall in the field of view, but it
is not visible (as SNe are typically only bright at UV wavelengths for
some weeks after explosion). We searched the CFHT, Gemini, Subaru,
ING, TNG and NOAO archives, but no images of SN 2001ci were found. We
have hence not considered SN 2001ci any further.

\subsection{2002ap}

SN 2002ap was discovered in M74 by amateur astronomers \citep{Nak02},
and rapidly classified as a type Ic SN \citep{Mei02}. SN 2002ap
was extremely energetic, and bore a striking spectroscopic similarity
to the ``hypernovae'' 1997ef and 1998bw
\citep{2002ApJ...572L..61M,Gal02}. The explosion is thought to be
moderately asymmetric, both from early time spectropolarimetry
\citep{Kaw02,Wan03} and modelling of the late time nebular spectra
\citep{Maz07}. 

The first attempt to detect the progenitor of SN 2002ap was made by
\citet{Sma02}, who did not find a progenitor in ground-based
pre-explosion imaging. A subsequent study by \citet{Cro08} used even
deeper pre-explosion images, which still did not yield a progenitor
detection. The limits for SN 2002ap remain the deepest limits for a
hydrogen-deficient SN to date. The 5$\sigma$ limiting magnitudes found
by Crockett et al. are B= 26.0$\pm$0.2 mag, and R=24.9$\pm$0.2
mag. The limits from \citet{Sma02} are much shallower, with the
exception of their U-band image, which is of interest for constraining
a hot progenitor, and had a 3$\sigma$ limit of 21.5 mag. 
\cite{2005MNRAS.359..906H}
 considered the various measurements of the distance towards the
host galaxy for SN 2002ap, M74, and adopted an average value of 9.3
$\pm$ 1.8 Mpc, which we also adopt \citep[as in][]{Cro08}. 
 The total extinction (Milky Way
and host galaxy) towards SN 2002ap is given by \citet{Tak02} as
E(B-V)=0.09$\pm$0.01, from a high-resolution spectrum of the Na{\sc
  i}D lines.
         

\subsection{2003jg}

SN 2003ig was discovered by LOTOSS \citep{Gra03}, and classified by
\citet{Fil03} as a Type Ic SN approximately a week after maximum
light. The SN exploded in NGC 2997, which has a Tully-Fisher distance
of 12.2 $\pm$0.9 Mpc \citep{Hes09}, which is in good agreement with
the recessional velocity based distance of 12.7 Mpc. The foreground
extinction towards NGC 2997 from \citet{Sch98} is E(B-V)=0.109 mag.

Post explosion images of SN 2003jg were obtained in several filters
with HST+ACS/HRC on 2003 Nov 18. Pre-explosion WFPC2 images of NGC
2997 as detailed in Table \ref{tab_obsdata} were used to search for
the progenitor. The images in each filter consisted of a cr-split
pair, these were combined with the {\sc crrej} task in {\sc iraf} to
reject cosmic rays. The pre-explosion WFPC2 F450W and F814W images
were taken at the same pointing, with the SN position falling on the
WF3 chip.  To accurately identify the SN position in these, the F814W
image was aligned to the post explosion ACS/HRC F814W image. The SN
position in pre-explosion F606W (on the PC chip) image was determined
by aligning it to the post-explosion ACS/HRC F555W image.

20 sources were identified common to both the F814W ACS and F814W
WFPC2 images, and used to derive a general transformation with an rms
error of 17 mas. The SN position was measured in the ACS F814W image
(j8qg10021\_drz) to be (704.530, 259.953), with an uncertainty of 0.3
mas, which was transformed to the pixel coordinates of the WFPC2 F814W
image (u6ea3803r\_c0f), where it was found to be 246.10, 164.15, with
an uncertainty of 0.2 pixels

The pre- and post-explosion F606W WFPC and F555W ACS images were
aligned with an rms error of 16 mas from a general transformation to
29 common sources; the SN position was measured in the post explosion
image to lie at pixel coordinates 704.844, 259.740 in j8qg10050\_drz
with an rms error of 1 mas. The SN position was transformed to the
pre-explosion WFPC2 image, where it was found to have pixel
coordinates on the PC chip of 334.99, 166.94 in u29r1301t\_c0f, with
a total uncertainty of 0.3 pixels.

\begin{figure*}
\subfigure[Pre-explosion WFPC2/WF3 F450W image]{
\includegraphics[width=41.5mm,angle=0]{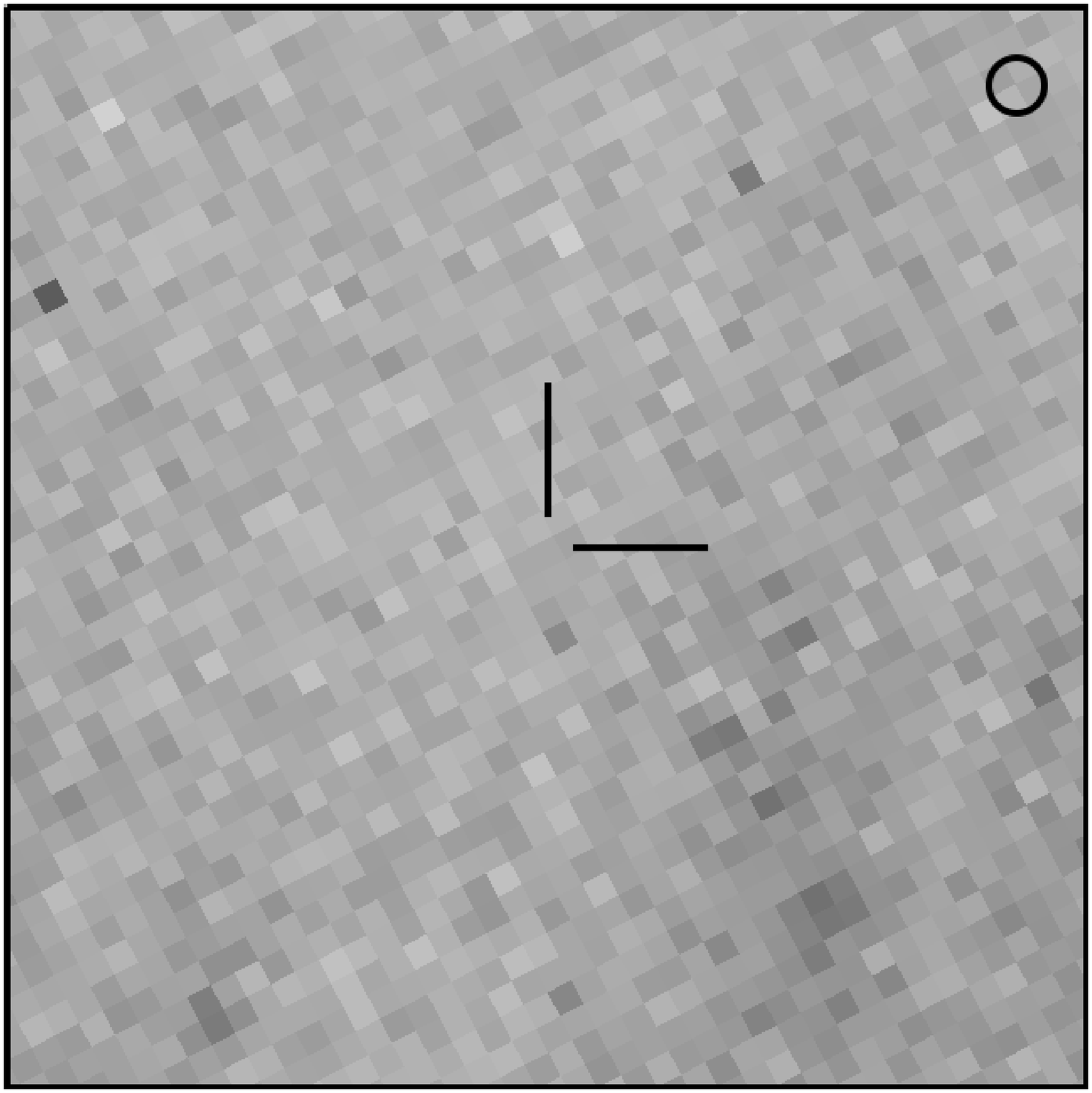}
\label{fig2:subfig1}
}
\subfigure[Pre-explosion WFPC2/PC F606W image]{
\includegraphics[width=41.5mm,angle=0]{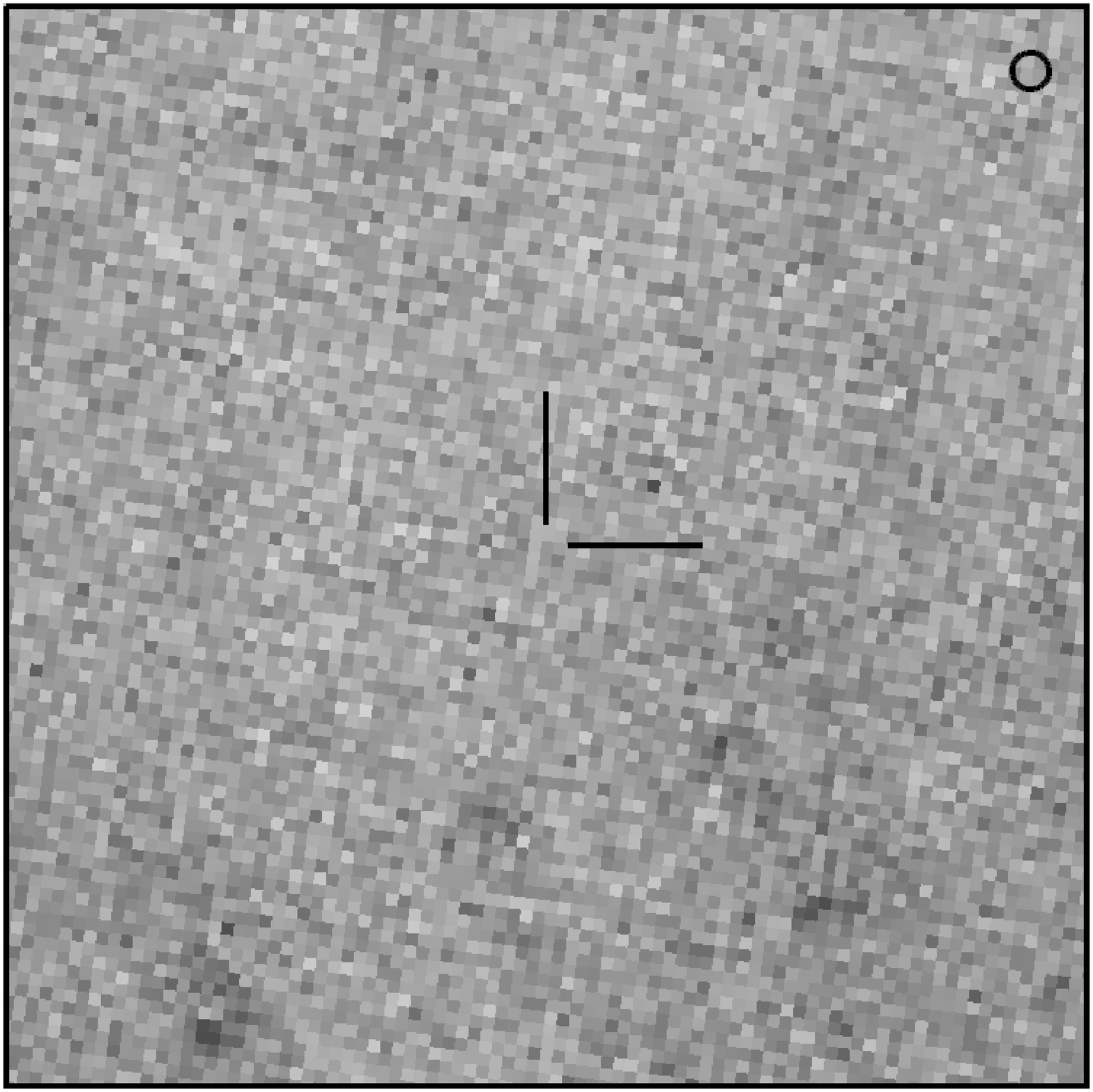}
\label{fig2:subfig2}
}
\subfigure[Pre-explosion WFPC2/WF3 F814W image]{
\includegraphics[width=41.5mm,angle=0]{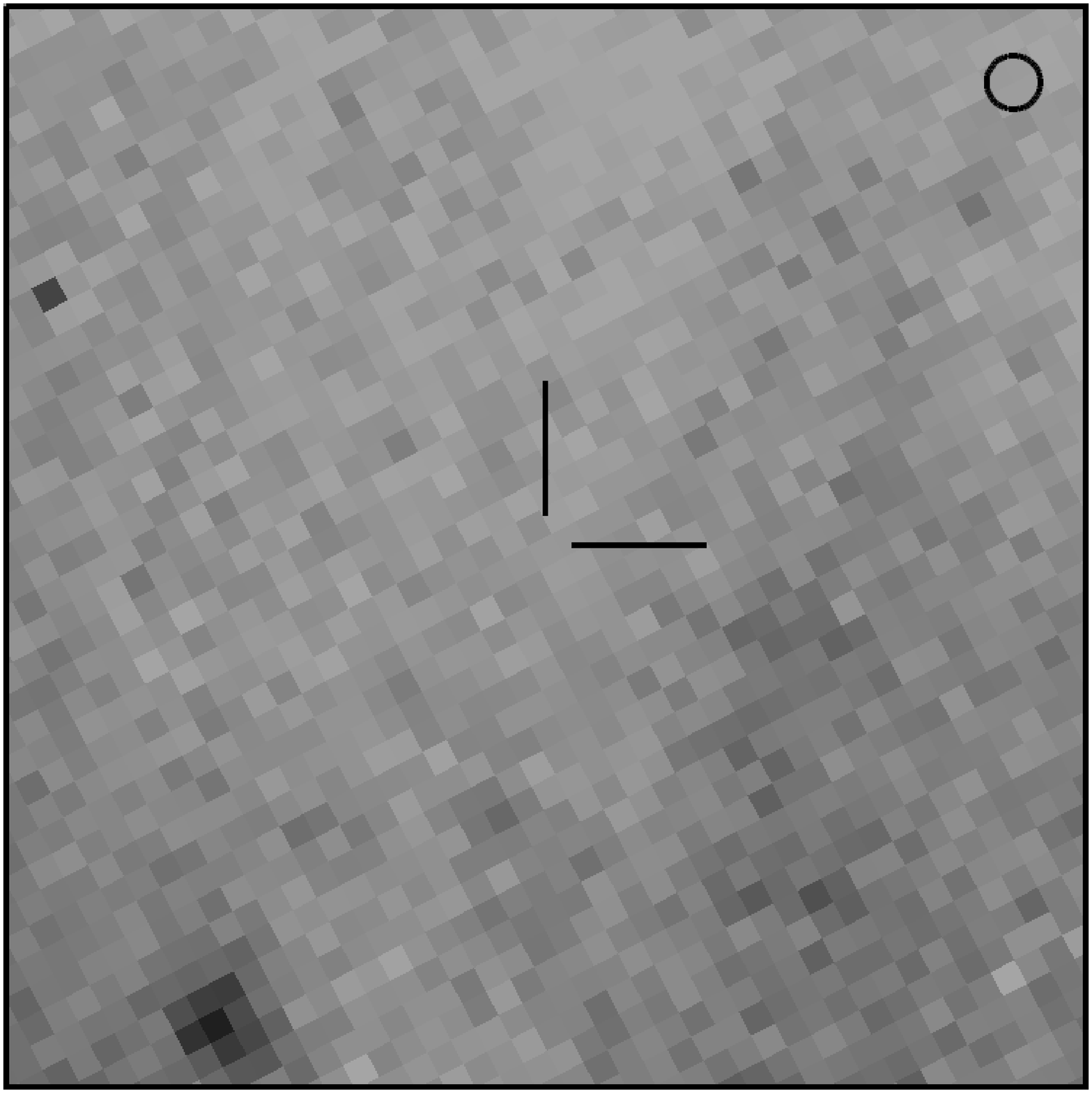}
\label{fig2:subfig3}
}
\subfigure[Post-explosion ACS/HRC F814W image]{
\includegraphics[width=41.5mm,angle=0]{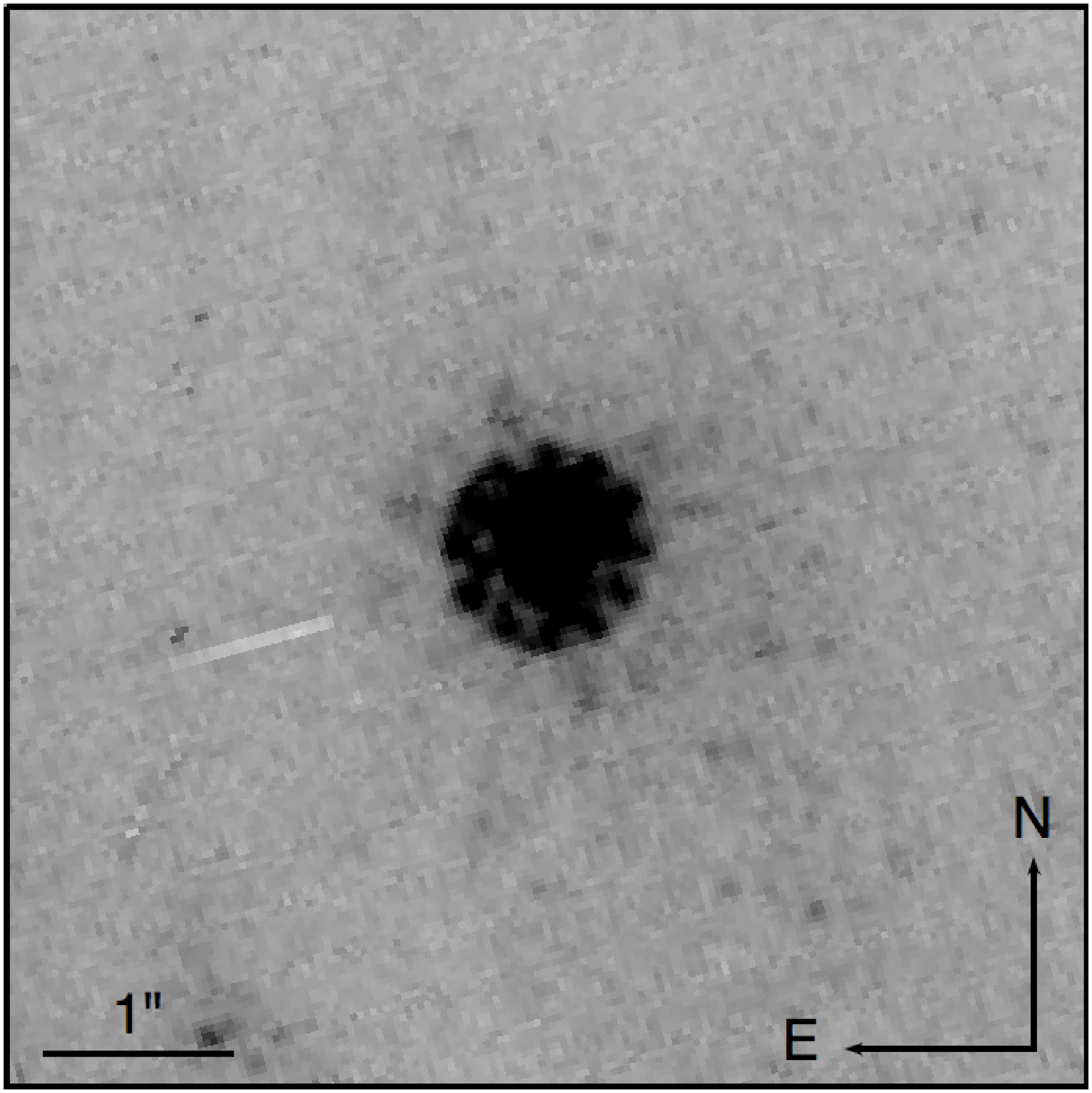}
\label{fig2:subfig4}
}
\caption{Pre-explosion HST+WFPC2 images of the site of SN 2003jg, plus
  a post-explosion HST+ACS image. Scale and orientation for each panel
  is identical to that indicated in the post-explosion image.  The
  circle in the upper right corner of each panel has a radius equal to
  5 times the rms error in the position of the supernova in that
  image.}
\label{fig_SN2003ig}
\end {figure*}

Inspection of the transformed SN position in the pre-explosion frames
did not reveal any coincident source, as can be seen in
Fig. \ref{fig_SN2003ig}. {\sc hstphot} was run with a 3$\sigma$
detection threshold on the pre-explosion images, but did not detect a
source coincident to the SN at this significance. We calculate
5$\sigma$ limiting magnitudes for the progenitor of SN 2003jg to be
F450W$>$24.63, F606W$>$24.63 and F814W$>$23.70 using the technique
described in \citet{Cro11}.

\subsection{2004gk}

SN 2004gk was discovered by \citet{2004IAUC.8446....1Q} and classified
as a Type Ic SN near maximum light. The host galaxy of SN 2004gk, IC
3311, is at an inclination angle of 90$\deg$ (from LEDA), rendering
any search for the progenitor optimistic at best. We note however that
\citet{Elm11} report a host galaxy extinction of $A_V$=0.372 mag from
low resolution spectroscopy, while from photometry the SN does not
appear to be heavily reddened. It is hence plausible that while IC
3311 is edge on to our line of sight, SN 2004gk may have exploded
relatively close to the edge of the galaxy. The galaxy has a negative
recessional velocity as it lies within the Virgo supercluster, however
the Tully-Fischer distance to the galaxy from LEDA is $\sim$20
Mpc. Unfortunately, there are no pre-explosion HST images of the
galaxy, while natural-seeing ground based imaging is not useful at
such distances. We have hence not considered SN 2004gk any further in
the following.

\subsection{2004gn}

SN 2004gn was discovered with the Katzman Automatic Imaging Telescope
(KAIT) as part of the Lick Observatory Supernova Search (LOSS) close
to the nucleus of NGC 4527 \citep{Li04}.  The spectral classification
of SN2004gn was not reported in an IAU CBET. Private communication
from the KAIT team (B. Cenko) confirms that it is a H-deficient SN and
a Ibc classification is reasonable.
\footnote{http://astro.berkeley.edu/bait/public\_html/2004/sn2004gn.html}.  A Nearby Supernova Factory spectrum
\footnote{http://www.rochesterastronomy.org/sn2004/sn2004gn.jpg} shows
it to be a Ibc similar to SN1990B at 90 days, which is a Ib. As the SN
was discovered late, we cannot rule out that there was residual
hydrogen at early times that has disappeared, but with the lack of
early data we assume a classification of Ibc. We found very late time
spectroscopy of the SN location from the WHT in the ING archive,
however no flux from the SN was present and so this was of no use for
this work.

Besides SN 2004gn, NGC 4527 was also host to the peculiar Type Ia SN
1991T. Subsequently, the galaxy has been the subject of numerous
Cepheid studies with HST to obtain an independent measure of the
distance. The most recent of these \citep{Sah06} found a distance of
14.2$\pm$1.3 Mpc towards NGC 4527. SN 1991T itself implies a distance
of 14.1 Mpc from fits to the multi-colour light curve
\citep{Jha07}. We have adopted the average of these two values
(14.2$\pm$1.3 Mpc) as the distance towards the host. The extinction
towards SN 2004gn is poorly constrained, as we have no spectra, or
information on phase to allow us to compare the magnitude to other
Type Ibc SNe. Hence we adopt the foreground extinction from NED,
A$_V$=0.07 mag, as a lower limit to the extinction.

Broad-band pre-explosion images for SN 2004gn (as listed in Table
\ref{tab_obsdata}) consist of two 300s WFPC2 F606W images, where the SN
falls on the WF3 chip. Unfortunately no post explosion images of the
SN were discovered during our archival search. As an alternative to a
direct alignment between pre- and post- explosion frames, we attempted
to use three isolated SDSS point sources to register the combined
WFPC2 WF3 image to the celestial WCS. Attempting to solve this with
only three sources means we are over fitting the data, and it does not
make sense to quote an rms error in the alignment. However, by
inspection of the positions of other SDSS sources (which were not used
in determining the transformation, as they are extended), we believe
that the accuracy of the registration is $\lesssim$1\arcsec.

\begin{figure}
\includegraphics[angle=0, width=84mm]{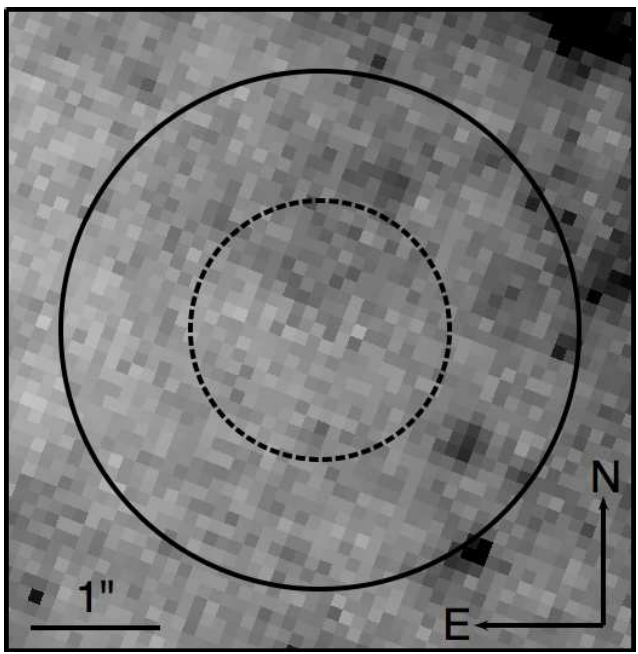}
\caption{Pre-explosion HST WFPC2 image of the site of SN 2004gn, taken with the F606W filter. As no post-explosion images were available, we show a 1\arcsec (dashed line) and 2\arcsec (solid line) radius around the published coordinates for the SN.}
\label{fig_SN2004gn}
\end{figure}

The RA and Dec of SN 2004gn as reported on the IAU webpages correspond
to pixel coordinates of 77.8, 204.0 on the WF3 chip (using the new WCS
solution). We ran {\sc hstphot} with a detection threshold of
3$\sigma$. Within a 1\arcsec radius of the reported SN coordinates
(shown in Fig. \ref{fig_SN2004gn}), we find a single source detected
above the 5$\sigma$ level, at a magnitude of F606W=25.4. Six other
sources are detected above 3$\sigma$, but below 5$\sigma$, and with
magnitudes between 25.4$<$F606W$<$25.9. Within the larger annulus,
seven sources are detected above the 5$\sigma$ level, the brightest of
which has a magnitude of F606W=24.2. The brightest source has a large
$\chi^2$ value, and a sharpness parameter which is more negative than
would be expected for a single star. In fact, all of the sources
detected by {\sc hstphot} have negative sharpness, suggesting that
these are not stellar, but rather that {\sc hstphot} is detecting
unresolved background structure. Nevertheless, we adopt the magnitude
of the brightest source detected by {\sc hstphot} as our limit for the
progenitor magnitude, i.e. F606W$>$25.4. The F658N images were
examined, but no H$\alpha$ excess was seen at the SN position.

\subsection{2004gt}

\citet{Mon04} discovered SN 2004gt in NGC 4038; the SN was
spectroscopically classified as a Type Ic SN. \citet{Kin04} suggested
that it was difficult to distinguish between the Ib and Ic subclasses
therefore we take the conservative approach and label the SNe as Type
Ibc. NGC 4038 is part of a well the studied interacting galaxy pair
Arp 244, better known as the Antennae galaxies. Despite extensive
observations of the Antennae, their distance is still a matter of some
debate in the literature. \citet{Sav04} derived a distance of
13.8$\pm$1.7 Mpc from the tip of the red giant branch (TRGB) using HST
WFPC2 imaging. This value was revised by \citet{Sav08} to 13.3$\pm$1.0
Mpc, based on deeper images from HST+ACS. The TRGB distance is
considerably closer than that expected given the recessional velocity
of Arp 244, which implies a distance of $\sim$20 Mpc. The discovery of
the Type Ia SN 2007sr in NGC 4038 provided an independent measurement
of the distance from a fit to the SN lightcurve, yielding a value of
22.3$\pm$2.8 Mpc, in disagreement with the TRGB estimate
\citep{Sch08}. Schweizer et al. suggest that Saviane et al. have
misidentified the TRGB, and present a re-analysis of the HST data used
by the latter which brings the TRGB distance into accord with that
from SN 2007sr. We follow the suggestion of Schweizer et al. in
adopting 22$\pm$3 Mpc as the distance for NGC 4038, based on an
average of the values from the recessional velocity, re-calibrated
TRGB, and SN 2007sr. The extinction towards SN 2004gt has been taken
as E(B-V)=0.07$\pm$0.01, following the lead of \citet{Mau05}.

Both \citet{Mau05} and \citet{2005ApJ...630L..29G} presented limits on
the progenitor of SN 2004gt from pre-explosion HST images. Maund et
al. found limits of F336W $>$ 23.04, F439W $>$ 24.56, F555W $>$ 25.86
and F814W $>$ 24.43. The limits found by Gal-Yam et al. are similar,
but as their values are quoted in $UBVI$ rather that the HST flight
system these cannot be compared directly. Gal-Yam et al. also present
a limit from a pre-explosion STIS image of $>$23.4 for the far-UV.

Using this new distance and extinction we determine upper limits for
the absolute magnitudes for the progenitor of SN2004gt in
Table\,\ref{obsprogenitors}.


\subsection{2005V}

The host galaxy of SN 2005V, NGC 2146, is a starburst galaxy, and is
thought to have a large population of recently formed massive
stars. However, the absence of an overabundance of supernova remnants
may indicate that the starburst is at an earlier stage than similar
galaxies such as M82 \citep{Tar00}. The kinematic and Tully-Fischer
estimates for the distance towards NGC 2146 agree well (17.2 and 16.9
Mpc respectively), however the sosies estimate is significantly larger
at 27.7 Mpc. As NGC 2146 appears to have a slightly irregular
morphology, it is possible that the sosies method is not a reliable
measure in this case. We have hence discarded this measure, and
adopted the mean of the kinematic and Tully-Fischer distances.

\citet{Mat05} discovered SN 2005V in the near infrared, although the
early time J-H and H-K colours of the SN do not support very high
levels of extinction. Mattila et al. find J-H=0.1 mag, H-K =0.2 mag
for SN 2005V, which is indistinguishable from the the typical J-H and
H-K colours of Ibc SNe \citep{Hun09}. \citet{2005IAUC.8474....3T}
classified the SN as a Type Ibc, but with a strong Na{\sc i}D
absorption (EW = 5.5 \AA) and a red continuum which does indicate
significant reddening. Unfortunately there is no published photometry
of SN 2005V, so we cannot compare the optical colours and absolute
magnitude to other Type Ibc SNe. Using the relation of \citet{Tur03}
between NaD absorption and reddening, we infer E(B-V)$\sim$0.9,
however we are somewhat sceptical of the veracity of this estimate due to the large
scatter in the relation \citep{Poz11}. Nonetheless we adopt E(B-V) as
the conservative option, as an {\it over}estimate of A$_V$ will only
serve to make our limits {\it less} restrictive. On a more qualitative
level, we also note that the SN is close to the nucleus of the galaxy,
that a large number of dust lanes are seen in the HST images, and that
the SN position is offset by $\sim$0.15 \arcsec from one of these
lanes.

\begin{figure*}
\subfigure[Pre-explosion WFPC2/WF3 $F450W$ image]{
\includegraphics[width=41.5mm,angle=0]{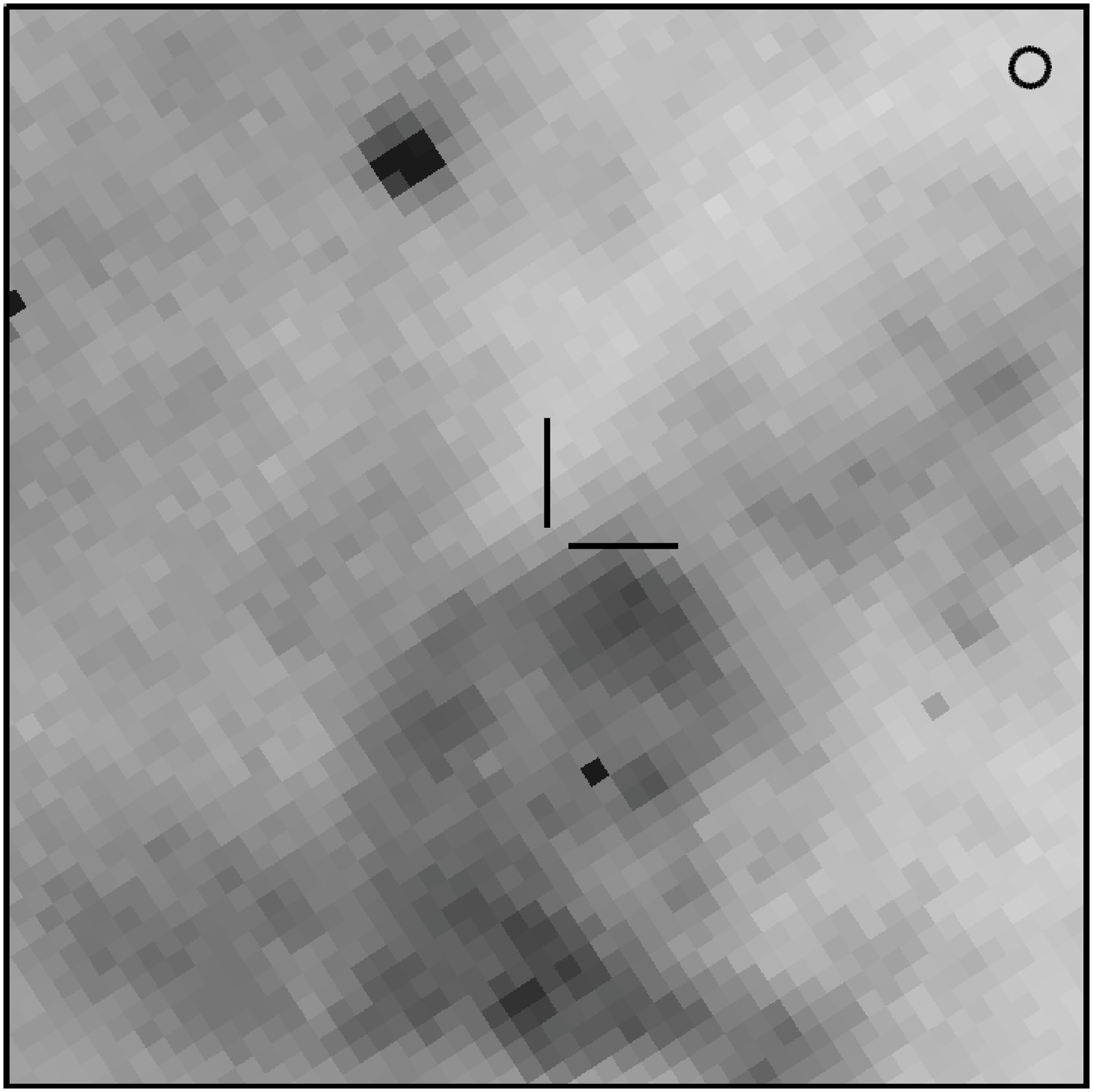}
\label{fig3:subfig1}
}
\subfigure[Pre-explosion WFPC2/PC $F606W$ image]{
\includegraphics[width=41.5mm,angle=0]{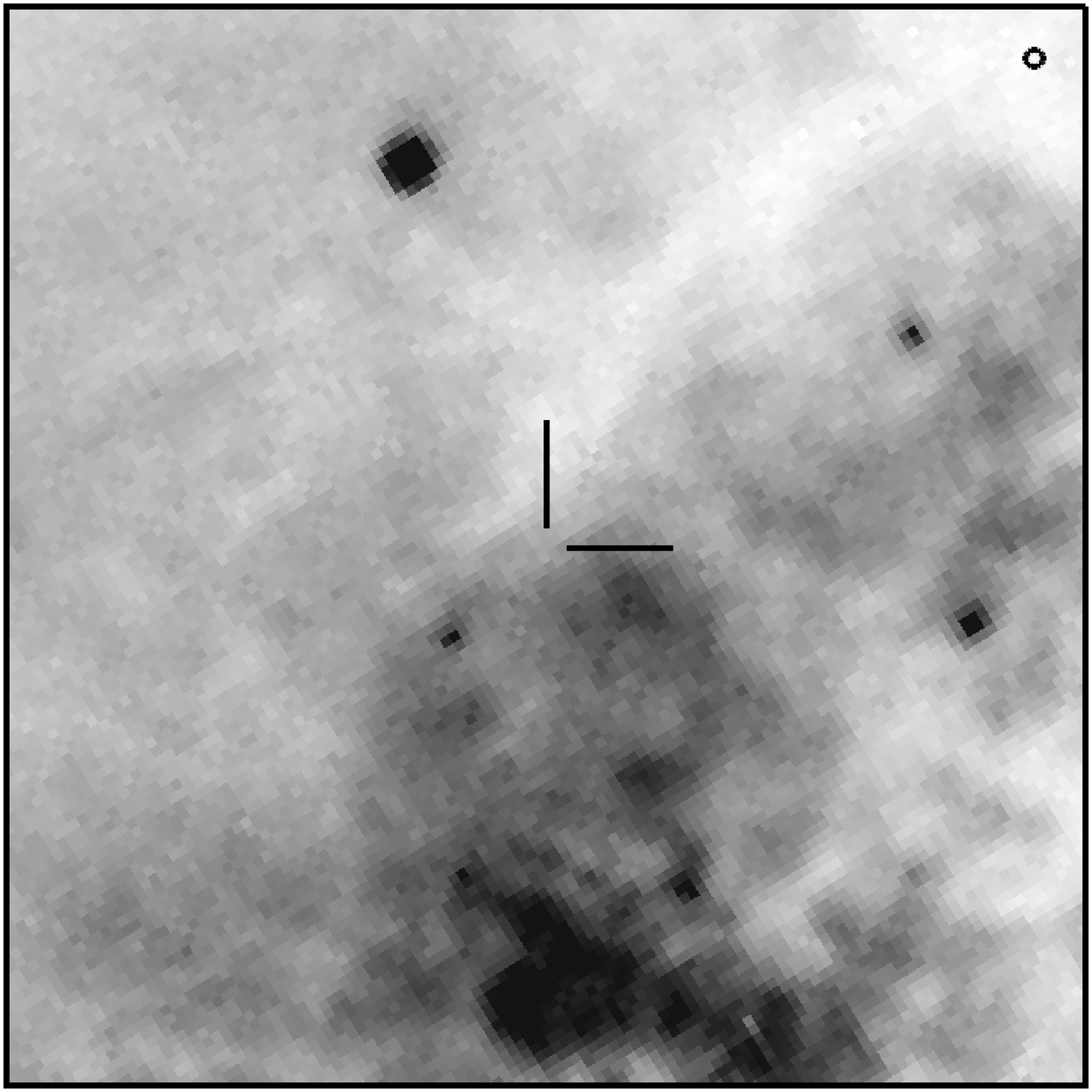}
\label{fig3:subfig2}
}
\subfigure[Pre-explosion WFPC2/WF3 $F814W$ image]{
\includegraphics[width=41.5mm,angle=0]{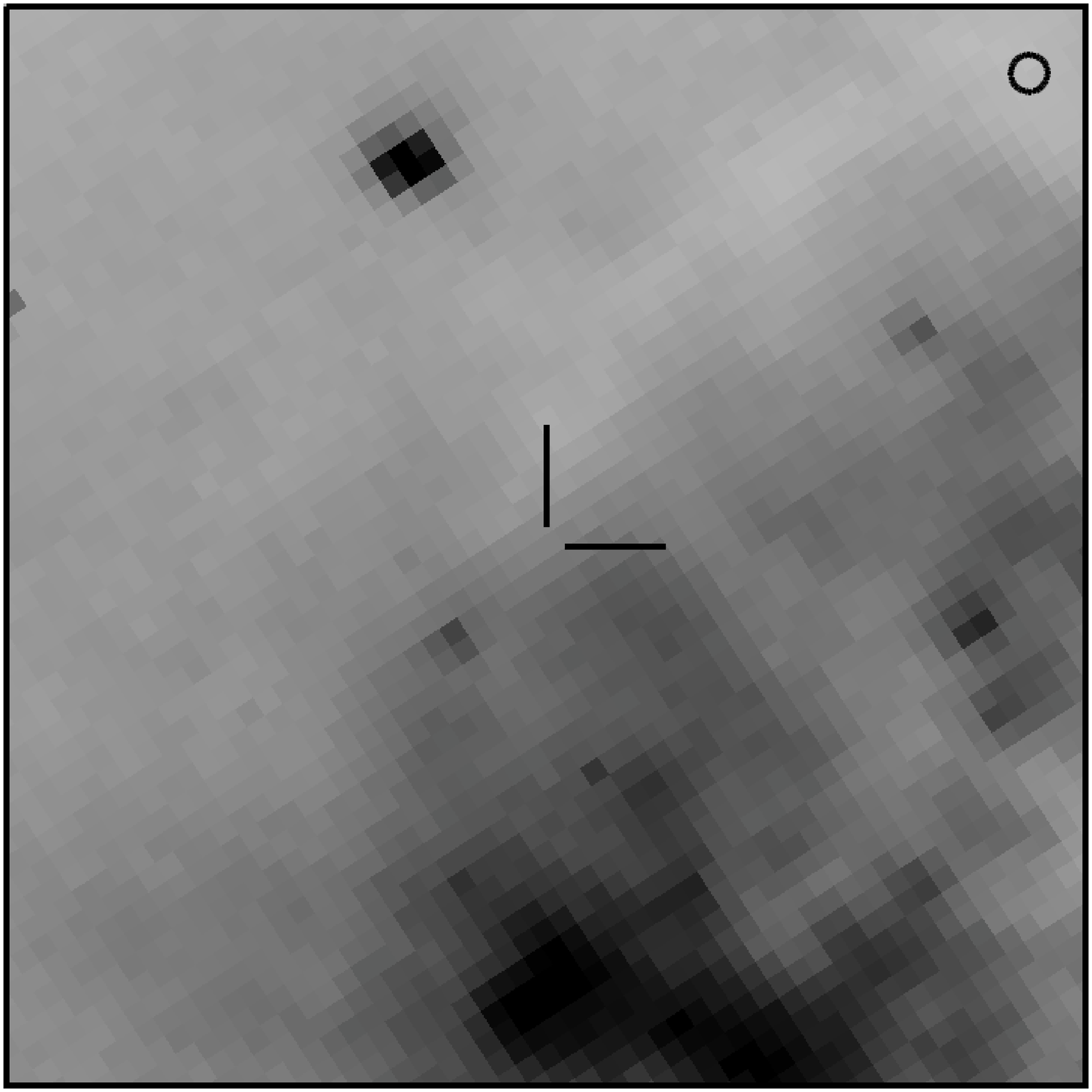}
\label{fig3:subfig3}
}
\subfigure[Post-explosion ACS/HRC $F814W$ image]{
\includegraphics[width=41.5mm,angle=0]{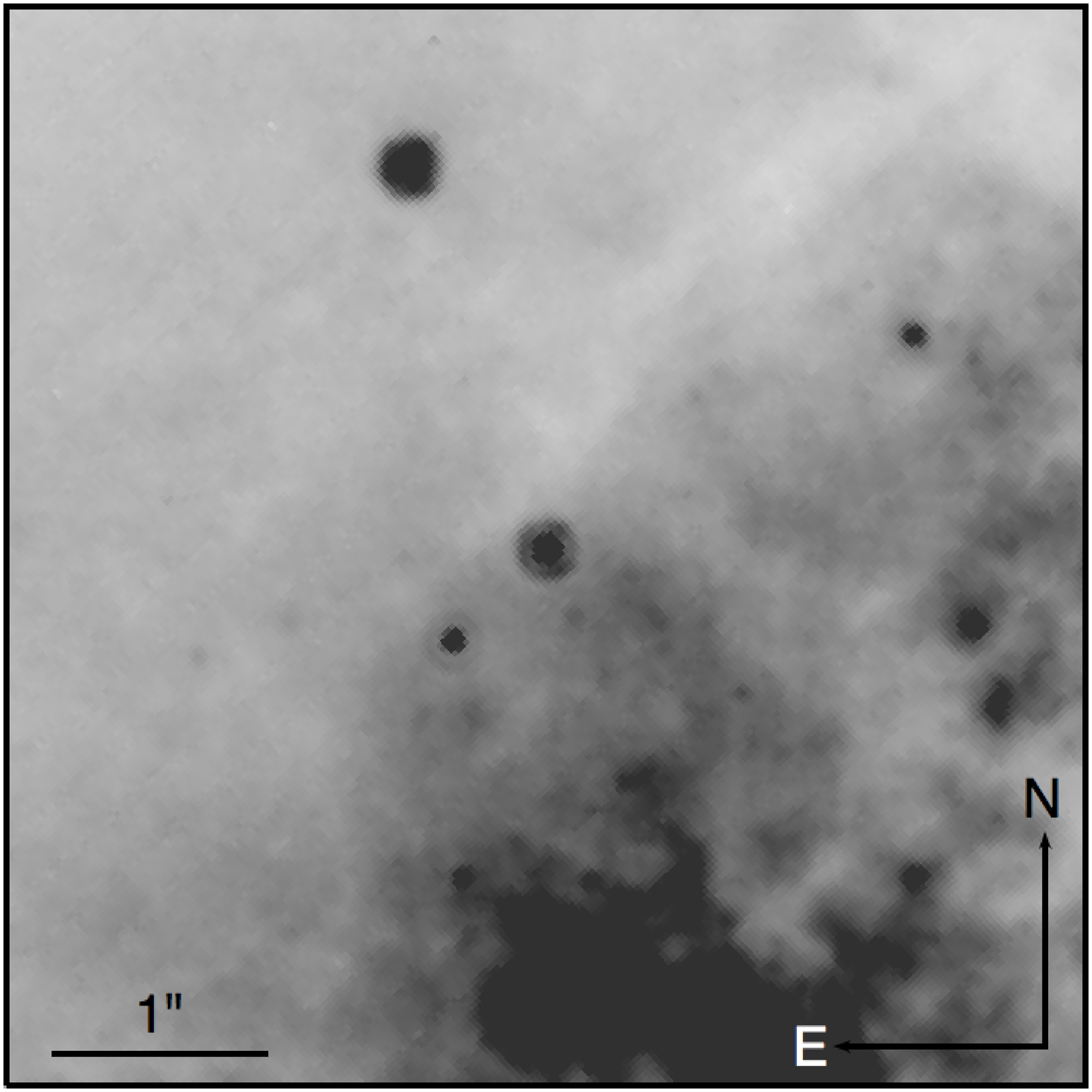}
\label{fig3:subfig4}
}
\caption{Pre-explosion HST+WFPC2 images of the site of SN 2005V, plus
  a post-explosion HST+ACS image. Each panel covers a 5\arcsec
  $\times$5\arcsec region. Scale and orientation for each panel is
  identical to that indicated in the post-explosion image.}
\label{fig_SN2005V}
\end {figure*}

Pre-explosion WFPC2 images as listed in Table \ref{tab_obsdata} were
downloaded from the HST archive. Cr-split pairs were combined with
{\sc crrej} to remove cosmic rays. The SN position lies on the PC chip
in the F606W image, and on the WF3 chip in the F450W and F814W filter
images, as shown in Fig. \ref{fig_SN2005V}. To identity the SN
position in the pre-explosion images, we used images of the SN
obtained with the ACS/HRC on 2005-08-08.

The F606W pre-explosion was aligned to the F555W post-explosion
image. 26 sources were used to fit a general transformation with an
rms error of 0.164 WFPC2/PC pixels, corresponding to 8 mas. The SN
position was measured in j90n04020\_drz to be 578.770,609.971 with an
error of $\ll$1 mas. The SN position was transformed to the
pre-explosion image, where it was found to have pixel coordinates
331.27,411.27 in u67n1801r\_c0f.fits, with an uncertainty of 0.16
pixels. {\sc hstphot} did not detect a coincident source at the
3$\sigma$ level, and we calculated the limiting $5\sigma$ magnitude as
before to be F606W$>$24.76.

The F814W pre-explosion image was aligned to the post-explosion F814W
image with 17 sources, for an rms error of 0.17 WFPC2/WF pixels, or 17
mas. The SN position was measured in the ACS/HRC F814W image to an
accuracy of 1.5 mas, and transformed to the pre-explosion WFPC2 F814W
image. We derived transformed coordinates of 140.19, 94.01 for the SN
in u6ea2803r\_c0f. No source was evident at the SN location, and {\sc
  hstphot} detected no source at the $3\sigma$ level or above in
either the F450W or F814W images. We calculate 5$\sigma$ limiting
magnitudes for the progenitor of SN 2005V of F814W$>$22.80,
F450W$>$24.05.

The ACS pre-explosion images were also examined. The F814W image
showed no source coincident with the SN, and as the limit is
comparable to that from the WFPC2 F814W image, this was not considered
any further. The F658N image (narrow-band H$\alpha$) showed flux at
the SN position, but as there was no corresponding continuum
observations, it was unclear whether this flux comes from H$\alpha$
emission or was simply from the high background levels. The NICMOS
data were taken with the NIC3 camera, which has a pixel scale of
0.2\arcsec. The narrow band images were not examined in details, while
the broad band F160W image was too shallow to be of use, especially
for constraining a compact (and presumably hot) progenitor.

\subsection{2005at}

SN 2005at exploded in NGC 6744, at a distance of $14.1\pm2.8$
Mpc. Unfortunately the SN suffers from a high level of extinction
(A$_V\sim2.3\pm0.3$, E. Kankare, private communication), while the
pre-explosion data consists of ground-based imaging (from
VLT+FORS,VIMOS,ISAAC and the ESO 2.2m+WFI) rather than from HST. Using
the ESO exposure time
calculators\footnote{http://www.eso.org/observing/etc/} we estimated
5$\sigma$ limiting magnitudes for the pre-explosion images, and took
these together with the distance modulus and extinction A$_V\sim2.3$
to estimate the absolute magnitude reached be the pre-explosion
data. As the absolute magnitude reached is between -9 and -11 mag,
which is between 1.5 and 3.5 mags brighter than the most luminous WR
star in Fig. \ref{WRprogenitorcolours1}, these are of no use for
constraining the progenitor. Furthermore, late time images of NGC 6744
obtained in 2012, when subtracted from the pre-explosion images show
no sign of a progenitor disappearance (E. Kankare, private
communication). We have hence not considered SN 2005at any further in
this work.



\subsection{2007gr}

SN 2007gr is a Type Ic SN which exploded in NGC 1058; the SN was first
discovered by \citet{Mad07}, and spectroscopically classified by
\citet{Cho07}. The photometric evolution of SN 2007gr was similar to
that of SN 2002ap \citep{Hun09}, but with marked spectroscopic
differences. Where spectra of SN 2002ap resembled those of the
broad-lined Ic 1998bw, SN 2007gr is spectroscopically similar to a
prototypical Type Ic such as SN 1994I. The spectra of SN 2007gr were
also notable for the presence of strong carbon absorption
\citep{2008ApJ...673L.155V}. Modelling of late time nebular spectra
for SN 2007gr \citep{Maz10} suggests an ejected oxygen mas of
0.8\msun, which implies that the CO-core that exploded resulted
from a star with ZAMS mass $\sim$15\msun. Such a progenitor would not have
been massive enough to lose its H and He envelopes through stellar
winds, and so must have been stripped as part of a binary system.

It was claimed by \citet{Par10} from radio interferometry that SN
2007gr was expanding at relativistic velocities. Together with the
absence of extremely high velocities in optical spectra, and the
polarization seen in spectropolarimetry \citep{Tan08}, Paragi et
al. explained the properties of SN 2007gr with a small amount of
material in a low-energy, bipolar jet (seen in radio), coupled with
mildly aspherical, non-relativistic ejecta. However, this
interpretation was questioned by \citet{Sod10}, who claimed from X-ray
and radio data that they could fit the evolution of the SN at these
wavelengths with a non-relativistic ejecta. 

\citet{crockett07gr} used pre-explosion HST and post-explosion Gemini
images to show that the SN exploded on the edge of what appears to be
either a small compact cluster or a bright supergiant star.  Assuming
it to be a cluster and fitting of the cluster age yielded uncertain
results; {\sc Starburst99} models give an age of $7\pm0.5$Myrs or 16
to 21 Myrs with progenitor masses of 28$\pm4M_{\odot}$ and 14 to
$11M_{\odot}$ respectively \citep{crockett07gr}.  Regardless of
whether the nearby source is a star or a cluster, Crockett et
al. (2008b) estimated detection limits for an unseen, point-like
progenitor; $F450W > 23.7$ and $F814W > 21.7$. These are not as deep as
typical HST WFPC2 images due to the proximity of the SN to this bright
object. Assuming $E(B-V)=0.08$ and a distance of 10.6~Mpc as adopted in
\citet{crockett07gr}, this corresponds to absolute magnitude limits
for the unseen progenitor of $F450W>-6.7$ and $F814W>-8.6$.



\subsection{2010br}

SN 2010br was discovered in NGC 4051 by amateur astronomers
\citep{Nev10} and spectroscopically classified as a Type Ibc SN
\citep{2010CBET.2245....2M}. NGC 4051 has a recessional velocity
distance of 12.7 Mpc (via NED), and a Tully-Fischer distance of 12.2
Mpc \citep{Tul09}. We adopt the average of these two values,
12.5$\pm$0.3 Mpc, as the distance to NGC 4051. The foreground
extinction towards SN 2010br is $A_V$=0.04 mag, from NED, which we
adopt as a lower limit to the extinction.

\begin{figure}
\subfigure[Pre-explosion WFPC2/PC $F606W$ image]{
\includegraphics[width=84mm,angle=0]{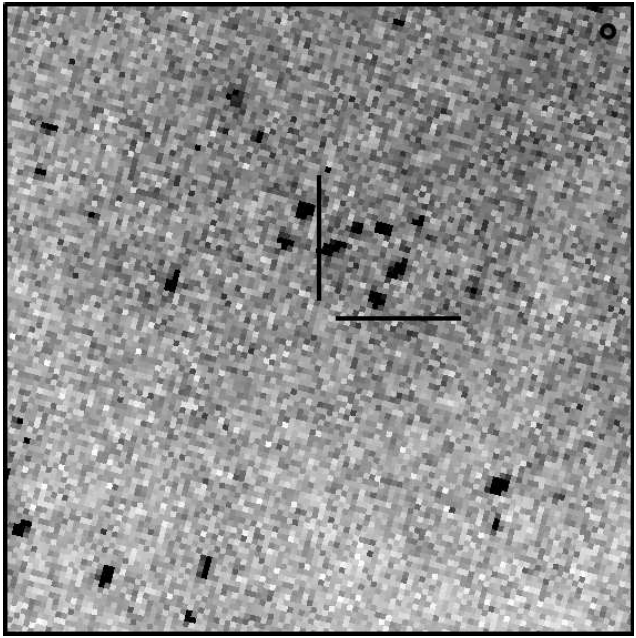}
}
\subfigure[Post-explosion WFC3 $F547M$ image]{
\includegraphics[width=84mm,angle=0]{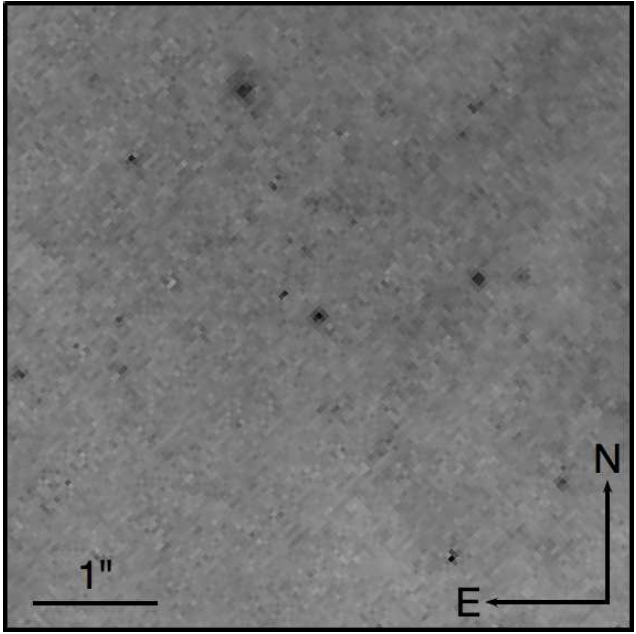}
}
\caption{Pre-explosion HST+WFPC2 image of the site of SN 2010br, plus
  post-explosion HST+WFC3 image. Each panel covers a 5\arcsec
  $\times$5\arcsec region. Scale and orientation for each panel is
  identical to that indicated in the post-explosion image, the circle
  in the upper right corresponds to 5 times the positional rms error.}
\label{fig_SN2010br}
\end{figure}

HST pre-explosion imaging consisted of a single WFPC2 F606W image
taken on 1994 June 06 (u2e64a01t\_c0f). Unfortunately the image was
not taken as a cr-split pair, so we have been careful to ensure that
there are no cosmic rays in the image at the position of SN
2010br. The SN fell on the PC chip, and so we aligned the PC chip only
to a 2340s post-explosion image taken in the F547M filter with WFC3 on
2010 July 17. Using 16 sources we obtained an rms error of 0.2 pixels
(10 mas) in a general fit. The SN position was measured in the
post-explosion image (ib5e04010\_drz) to be 2002.08, 2067.76, with an
rms error of $\ll$1 mas. The SN position was transformed to the
pre-explosion image, where it was found to be 670.77, 724.56 on the PC
chip. We show the images in Figure \ref{fig_SN2010br}.

No source was detected by {\sc hstphot} when it was ran on the
pre-explosion image with a 3$\sigma$ threshold. Using the same method
as previous, we calculate a limiting magnitude of $F606W>25.7$ for
the progenitor of SN 2010br.

\subsection{2011am}

\citet{Boc11} discovered SN 2011am in NGC
4219, and \citet{2011CBET.2667....1M} classified
the SN shortly after as a young Type Ib SN. Morrell et al. noted the
presence of strong Na{\sc i}D absorption at the redshift of the host,
suggesting significant extinction.  
\citet{Tul09} gives a
Tully-Fisher distance for the host of 21.0, while the kinematic
distance from NED (corrected for infall on Virgo, GA and Shapely) is
22.5 Mpc. We adopt the Tully-Fisher distance. 

Pre-explosion imaging of the site of SN 2011am consists of a 600s
WFPC2 exposure in the F606W filter, obtained on 1996 March 16. The SN
position lies close to the edge of the PC chip. We attempted to obtain
an adaptive optics image of SN 2011am with the VLT+NaCo, however the
correction obtained was unsatisfactory due to poor conditions at the
time of the observations. Instead, we used an $R$-band image obtained
of SN 2011am on 2011 Mar 27 with the NTT+EFOSC2 under good
($\sim$0.8\arcsec) seeing conditions to identify the progenitor. The
EFOSC2 image was aligned to a pre-explosion Hubble Legacy
Archive\footnote{\texttt{http://hla.stsci.edu/}} (HLA) mosaic
(u3321601b\_drz.fits) to an accuracy of 75 mas. The HLA mosaic was
then in turn aligned to the PC chip image, where the transformed SN
position was found to be 724.10,355.78, with an uncertainty of 76 mas.

\begin{figure}
\includegraphics[angle=0, width=84mm]{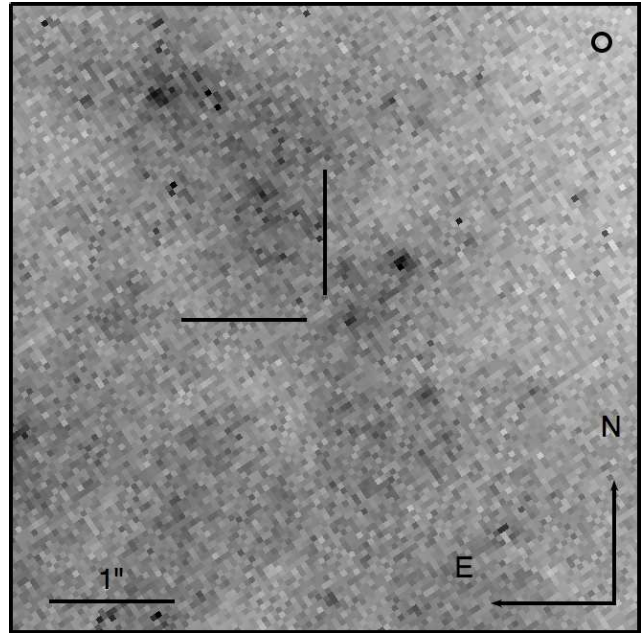}
\caption{ Pre-explosion HST+WFPC2 image of the site of SN 2011am. The panel covers a 5\arcsec$\times$5\arcsec\ region, scale and orientation are as indicated. The circle in the upper right corresponds to the positional rms error. 
}
\label{fig_SN2011am}
\end{figure}

As shown in Figure \ref{fig_SN2011am} SN 2011am is close (but probably
not coincident with) a source with $F606W=24.204\pm0.003$. This source
is also classified by {\sc hstphot} as extended. The formal 5$\sigma$
limit for the progenitor is 25.8 mag.  The extinction towards SN2011am
is uncertain but is almost certainly significant. The spectrum at
around -6d from Morrell et al. is significantly redder than that of
SN2007Y at a similar epoch \cite[-7 days
  from][]{2009ApJ...696..713S}. We de-reddened the spectrum of SN2007Y
by $E(B-V)=0.112$ as estimated by \cite{2009ApJ...696..713S}, and
de-reddened the spectrum of SN2011am to match the blue continuum of
SN2007Y, finding a very good match with $E(B-V)=0.6\pm0.1$. The
intermediate resolution spectrum of Morrell et al. nicely resolves the
Na\,D1 and D2 lines in the Milky Way and NGC4219. The Milky Way
components equivalent widths for D1 (5897\AA) and D2 (5891\AA) are
0.34\AA\ and 0.39\AA\ respectively. Using the
\cite{2012MNRAS.426.1465P} relation this suggests $E(B-V)=0.1\pm0.02$,
which is in reasonable agreement with the dust map line of sight
estimate of \cite{Sch98} $E(B-V)=0.13$.  However there is very strong
absorption in the ISM of NGC4219, giving equivalent widths for D1
(5897\AA) and D2 (5891\AA) of 0.86 and 1.26\AA. The
\cite{2012MNRAS.426.1465P} relation saturates at equivalent widths of
around 0.7\AA, hence can't be reliably used and we note that applying
it would give values of $E(B-V)=2.3 - 6.5$, both of which would give
unphysical blue spectral slopes and unrealistic absolute
magnitudes. Hence we adopt $E(B-V)=0.6\pm0.1$.  The corresponding
absolute magnitude is given in Table\,\ref{obsprogenitors}

\subsection{2011hp}

SN 2011hp exploded in the same galaxy as SN 2011am, and so we adopt
the same distance of 21.0\,Mpc \citep{Tul09}. The SN was first
discovered by \citet{2011CBET.2899....1M}, and classified by
\citet{Str11} as a Type Ic SN discovered around maximum light.
\citet{Fra11b} reported a limit for the progenitor of F606W $>$26.

\begin{figure}
\includegraphics[width=84mm,angle=0]{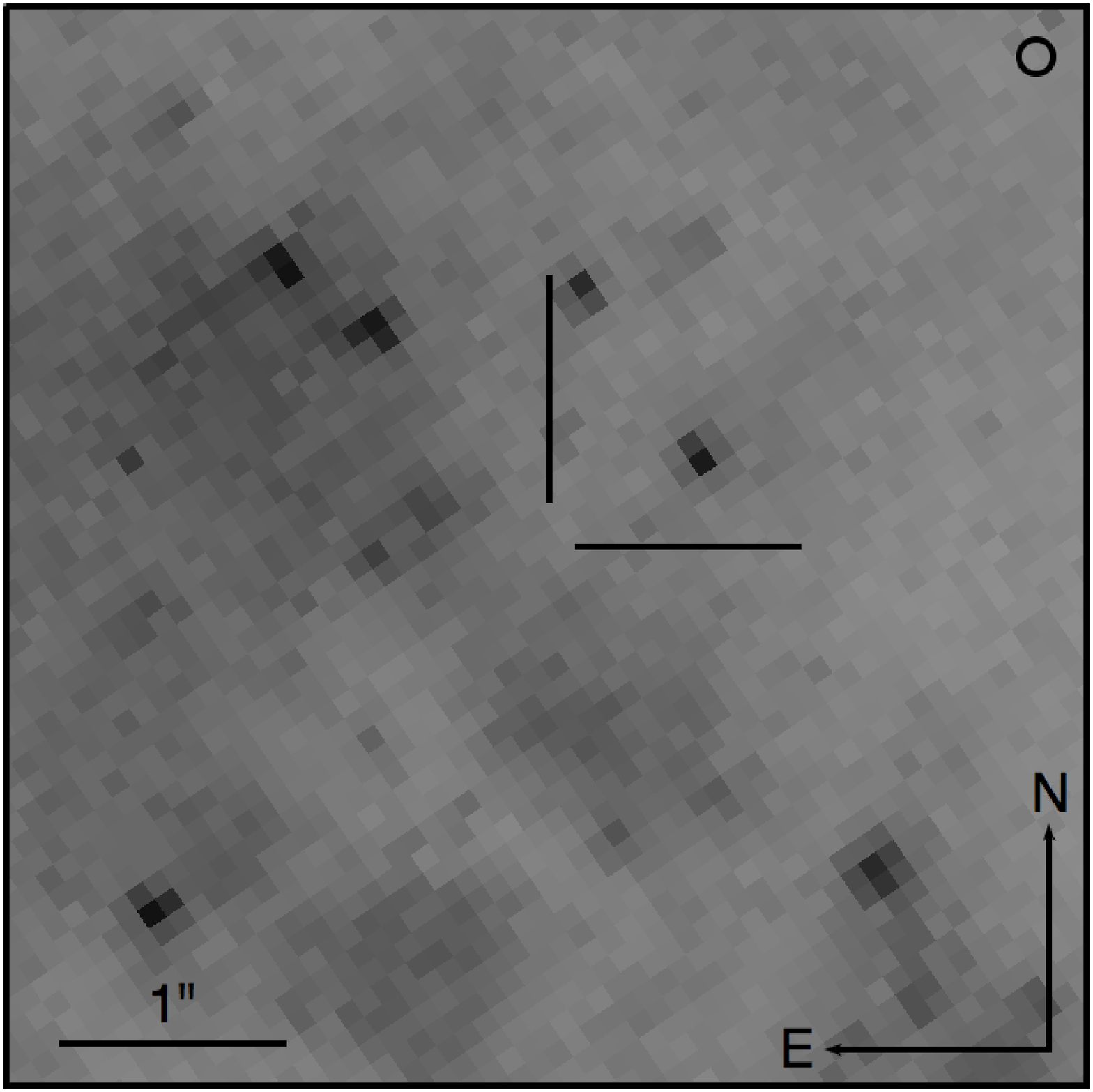}
\caption{Pre-explosion HST+WFPC2 image of the site of SN 2011hp. The
  panel covers a 5\arcsec $\times$5\arcsec region, scale and
  orientation are as indicated. The circle in the upper right
  corresponds to the positional rms error.}
\label{fig_11hw}
\end {figure}

The pre-explosion image for SN 2011hp is the same as that used for SN
2011am. Unfortunately we have no high resolution imaging for SN
2011hp, and so we are forced to rely on the same natural seeing image
as used by \citet{Fra11b} for a progenitor identification. We confirm
the position measured by Fraser et al. in the mosaic of all four WFPC2
detectors, where the SN falls on the WF4 chip, as shown in
Fig. \ref{fig_11hw}. However, we find a shallower limit than that
reported by Fraser et al., and calculate a new 5$\sigma$ limit for the
progenitor of SN 2011hp of $F606W>25.5$.

The spectra of 2011hp are also redder than normal, indicating
significant line of sight extinction similar to SN 2011am. We used a
spectrum from the ESO programme 184.D-1140 with EFOSC2 on the NTT
(3500-9000\AA, resolution 17.7\AA) taken at +15d after peak
(Elias-Rosa et al., in prep) to compare with the low extinction type
Ic 2007gr. The spectrum of 2011hp needs to be dereddened by
$E(B-V)\simeq0.5$ to match the spectral shape of 2007gr \citep[when it
  is dereddened by $E(B-V)=0.092$][]{}. Hence we adopt $E(B-V)=0.5$,
resulting in the absolute magnitude limit reported in
Table\,\ref{obsprogenitors}.

\subsection{2012au}
SN 2012au was identified in the Catalina Sky Survey, and subsequently
classified as a Type Ib SN (\citealp{2012CBET.3052....2S}). The host
galaxy, NGC 4790, has a recessional velocity distance (Virgo
corrected) of 21.8 Mpc, and a Tully-Fischer distance of 23.6 Mpc (both
from NED). We adopt the latter, corresponding to $\mu$=31.9, and a
foreground extinction of $A_V=0.13$. Pre-explosion imaging consists of
2$\times$160s images in each of $F450W$, $F606W$ and $F814W$, taken with
HST+WFPC2. We did not obtain an AO image of SN 2012au, however
\citet{2012ATel.3971....1V} set 3$\sigma$ upper limits of
$F450W>24.8$, $F606W>25.8$ and $F814W>25.1$ from an alignment with a
natural seeing image. Van Dyk et al. also identified a source
slightly outside the transformed SN position (offset by 0.2\arcsec) in
the F450W-filter image, with a magnitude of F450W=24.4. This
corresponds to an absolute magnitude in B of $-7.2$ mag. We have
re-estimated the limiting magnitudes in these images at the position of
the SN and find significantly brighter limiting magnitudes, due to the
very high sky background in this vicinity. We determine 5$\sigma$
limiting magnitudes of F814W$>23.4$, F606W$>24.5$ and F450W$>24.1$.

\section{Limits for the sample}
\label{s5}

Table \ref{obsprogenitors} contains the detection limits for twelve
hydrogen deficient core-collapse supernovae. We have not listed
metallicity information for these progenitors due to the lack of
homogeneous measurements for the host-galaxies. However all the hosts
are large spiral galaxies and therefore are expected to have
metallicities in the range between that of the Magellanic Clouds and
the Milky Way \citep[see discussion in][]{Sma09b}.  Distances in Table
\ref{obsprogenitors} are as discussed in the text. To gain insight
into the progenitors of Type Ibc SNe, we compare our observed limits
to a number of different plausible progenitor populations.

\begin{table*}
\caption{Properties and limits for the observed progenitors in our
   sample. The broad-band magnitudes with asterisks next to the value
   are in the standard Johnson-Cousins broad-band filters, the rest
   are \textsc{HST} filters with: $U$--$F336W$, $B$--$F450W$, $V$--$F555W$,
   $R$--$F606W$ and $I$--$F814W$.}
  \label{obsprogenitors}
   \begin{tabular}{@{}lccccccccccccc}
    \hline
    \hline
   		& SN   	&        		 	&               			&               	&     				&     				&    				&     				&   				\\
SN  		& Type 	& Galaxy 			& Dist(Mpc)		& E(B-V) 	& $M_U$ 			& $M_B$			& $M_V$			& $M_R$			& $M_I$			\\
\hline	
2000ew	& Ic		& NGC3810		& 18.2$\pm$3.3	& 0.04	& -				& -				& -				& -6.81 	& -				\\
2001B	& Ib		& IC391			& 25.5$\pm$2.5	& 0.14	& -				& -				& -8.25 	& -				& -				\\
2002ap	& Ic		& NGC628		& 9.3$\pm$1.8		& 0.09	& -8.85*		& -4.4*			& -				& -5.5*			& -				\\
2003jg	& Ic		& NGC2997		& 12.5$\pm$1.2 	& 0.11	& -				& -6.30 	& -				& -6.16 	& -6.99	\\
2004gn	& Ic		& NGC4527		& 14.2$\pm$1.3	& 0.07 	& -				& -				& -				& -5.55 	& -				\\
2004gt	& Ibc		& NGC4038		& 22$\pm$3	& 0.07	& -9.15 	&-7.44 	& -6.07 	& -				& -7.49*			\\
2005V	& Ibc		& NGC2146		& 17.1$\pm$3.2	& 0.90	& -				& -10.60 	& -				& -8.80 	& -9.93 	\\
2007gr	& Ic		& NGC1048		& 10.6$\pm$1.3	& 0.08    	&	                  	&   -6.7 			& -				& -				& -8.6	& -		\\
2010br	& Ibc	& NGC4051		& 12.7$\pm$2.0 	& 0.04 	& -				& -				& -				& -4.92 	& -				\\
2011am	& Ib		& NGC4219		& 24.5$\pm$5.6	& 0.60 	& -				& -				& -				& -7.44 	& -				\\
2011hp	& Ic		& NGC4219		& 24.5$\pm$5.6 	& 0.50      & -				& -				& -                           & -7.46 	& -				\\
2012au     & Ib        & NGC4790              & 23.6$\pm$2.0      & 0.04     &   -                         & -8.0                      & -                           & -7.5         & -8.6             \\
\hline	
\hline
\end{tabular}
\end{table*}

\subsection{Limits on progenitors from the observed WR star population}

A reasonable hypothesis is that the type Ibc SNe come from massive
Wolf Rayet stars. These are stars which will produce Fe-cores and have
atmospheres devoid of hydrogen and helium (the WN and WC+WO stars
respectively). The classical WR stars are thought to arise in very
massive stars that lose their envelopes through stellar winds,
potentially enhanced by binary interaction and stellar rotation.
These stars appear to be in clusters with turn-off masses that imply
lower limits of $>25$\msun\ in the Milky Way and Magellanic Clouds
\citep{2003ARA&A..41...15M,crowther2007}. Due to their high
temperatures and high mass-loss rates, the WR stellar radii, masses
and bolometric corrections are highly variable. With the strong
emission line spectra, this leads to a large range in the absolute
magnitudes of the stars.  Hence there is no easy way to set meaningful
luminosity or mass limits, as can be done for non-detections of II-P
progenitor stars \citep[e.g.][]{2003MNRAS.343..735S,Van03c} and we
need a simpler comparison method.  If we assume that this population
of stars are responsible for the type Ibc SNe that we see locally, we
can test if our non-detections are significant. In other words, what
is the probability that we have not detected a progenitor star simply
by chance? Of course this assumes that the WR stars that we observe
now do not change their optical fluxes significantly before
core-collapse, and we discuss this debatable point below.

We compare our limits to the observed magnitude distribution of WR
stars in the LMC. \cite{Mas02} presented a catalogue of $UBVR$
photometry for stars in the LMC, and used the WR catalogue of
\citet{1999A&AS..137..117B} to report broad-band magnitudes for
isolated WR stars. This is reasonably complete, and covers a large
range of WR luminosities and optical magnitudes.  As the catalogue of
Massey contains $BVR$ magnitudes for WR stars (whereas most of our
progenitor limits are in the HST $F450W$, $F555W$ and $F814W$ filters)
we have calculated HST -- Johnson-Cousins colours for each of the WR
spectral types, and used these to convert the $BVR$ magnitudes in the
\citet{Mas02} catalogue to the HST filter system.

We first fitted a function of form $Ax + B$ to the effective
temperatures of a sample of Galactic WR stars from \citet{Har06} of
spectral type WN$x$; where $A$ and $B$ are fitting coefficients. As
there were no WN1 stars in the sample of Harmann et al., we have
assumed a temperature of 180kK for this subtype. From this fit, we
have estimated a temperature for each of the WN spectral types, as
listed in Table \ref{wrtemptable}. Temperatures for the WO and WC
stars were taken from \citet{San12}. We then downloaded model WR
spectra from the Potsdam database
\citep{Har04},
\footnote{\texttt{http://www.astro.physik.uni-potsdam.de/$\sim$wrh/PoWR}}
appropriate to the temperature of each WR subtype. Models were
available for a range of transformed radii (effectively a range of
mass-loss rates) at each temperature; we have taken the models with
the largest and smallest transformed radii as the extreme cases. For
each model, we used the {\sc iraf.synphot} package to perform
synthetic photometry on the flux-calibrated model spectrum, and from
this calculate a colour as listed in Table \ref{wrtemptable}.

Along with spectral type, the colours of the WR stars depend strongly
on radius, as the emission lines which dominate WR spectra are
stronger in more compact models. The emission line strength will also
depend on the wind velocity of the WR star. As these are largely
unquantified uncertainties, we will simply adopt average $F450W$-$B$,
$F555W$-$V$ and $F606W$-$R$ colours for each of the WN, WC and WO
types, as listed in Table \ref{wrtemptable}.

\begin{table*}
\caption{Adopted temperature scale for WR types, together with colours
  from synthetic photometry of Potsdam models. For each colour, the
  two values listed are the colour as calculated from the model at
  that temperature with the smallest and largest transformed
  radii/mass-loss rates respectively.}
\label{wrtemptable}
\begin{tabular}{@{}lcrrrrr}
\hline
\hline 
WR Type	& Models (R$_{min}$|R$_{max}$)	& \teff (kK)	& $F450W$-$B$	& $F555W$-$V$	&$F606W$-$R$		& $F814W$-$I$		\\
\hline
WN9		& 03-10 | 03-04 	& 30		&0.026	| 0.042 	& -0.043	| -0.018 	& -0.047	| -0.090  		& -0.102 	| -0.090	\\											 		
WN8		& 04-15 | 04-04		& 35		&0.000 	| 0.041 	& -0.074	| -0.022 	& 0.016	| -0.086 		& -0.127 	| -0.085	\\	
WN7		& 05-16 | 05-04		& 40		&-0.002 	| 0.041 	& -0.081	| -0.030 	& 0.020 	| -0.090 		& -0.123 	| -0.081	\\			
WN6		& 07-18 | 07-04		& 50		&-0.006 	| 0.047 	& -0.090	| -0.032 	& 0.024 	| -0.092 		& -0.113 	| -0.083	\\				
WN5	 	& 09-20 | 09-04		& 60		&0.004 	| 0.051 	& -0.074	| -0.029 	& 0.035 	| -0.094 		& -0.119 	| -0.044	\\			
WN4		& 11-21 | 11-06		& 75		&0.003 	| 0.045 	& -0.088	| -0.038 	& 0.019 	| -0.096 		& -0.078 	| -0.042	\\
WN3		& 13-21 | 13-08		& 100	& -0.001 	| 0.039 	& -0.133	| -0.040 	& -0.01 	| -0.086  		& -0.079 	| -0.026	\\	
WN2		& 17-21 | 17-12		& 150	&-0.069 	|  0.037 	& -0.207	| -0.042 	& -0.03  	| -0.08 		& -0.139 	| -0.025	\\
WN1		& 18-21 | 18-13		& 180	&-0.083 	| 0.032 	& -0.226	| -0.043 	& -0.031	| -0.076 		& -0.096 	| -0.023	\\			
\hline				 			  	
WN		& mid-range		&		& -0.016			& -0.122			& -0.030				& -0.162			\\
\hline\hline
WC5		& 11-18 | 11-06		& 80		& 0.041	| 0.045  	& -0.059	| -0.034  	& 0.001	| -0.095 	 	& -0.879	| -0.875 	\\			
WC4		& 15-22 | 15-09		& 120	& 0.010	| 0.040 	& -0.099	| -0.042  	& 0.019	| -0.121 	 	& -0.712	| -0.648	\\	
\hline	\hline			 			  	
WC		& mid-range		&		& 0.028			& -0.071			& -0.051				& -0.764			\\
\hline							
WO3		& 19-26 | 19-24		& 200	& 0.004	| -0.064  	& -0.102	-0.204 	& 0.037 	| 0.168		& -0.727	| -0.667	\\				
\hline
WO		& mid-range		&		& -0.03			& -0.153			& 0.103				& -0.697			\\
\hline
\hline
\end{tabular}
\end{table*}

\begin{figure}
\includegraphics[width=0.47\textwidth,angle=0]{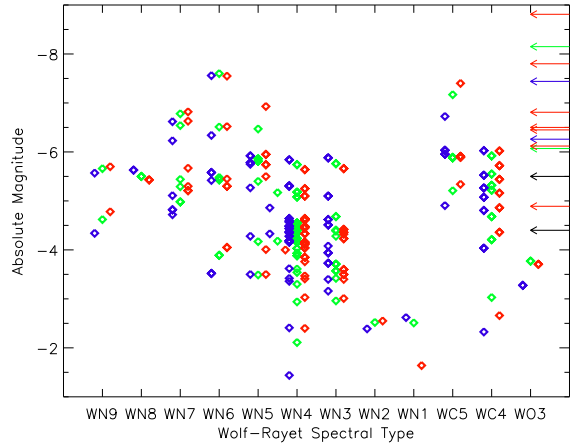}
\caption{Limiting magnitudes for the progenitors of Type Ibc SNe
  (indicated with arrows), compared to the observed magnitudes of LMC
  WR stars from \citet{Mas02}. WR star magnitudes have been converted
  to $F450W$, $F555W$ and $F606W$ (indicated by blue, green and red
  diamonds or arrows) as described in the text, and corrected for
  foreground extinction ($E(B-V)=0.080$) and the distance of the LMC
  ($\mu-18.50$ mag). All progenitor limits are in the HST filters,
  with the exception of those for SN 2002ap, which are in the
  Johnson-Cousins filter system.}
\label{WRprogenitorcolours1}
\end{figure}

We plot the observed sample of WR stars in Figure
\ref{WRprogenitorcolours1} along with the limits from our progenitor
sample in each of the $F450W$, $F555W$ and $F606W$ filters.


It is trivial to calculate the probability of not detecting any Type
Ibc progenitor thus far, assuming they are randomly drawn from the
population of WR stars shown in Fig. \ref{WRprogenitorcolours1}. For
each progenitor with a limit in B,V or R listed in Table
\ref{obsprogenitors}, we count the number of WR stars with a magnitude
brighter than this limit. For most of our Type Ibc SN sample, less
than $\sim$10 per cent of WR stars are brighter than our limit (and
hence can be ruled out). We can then calculate the probability of not
detecting {\it any} Type Ibc progenitors to date as $P=\Pi (1 -
F_{n})$, where $F_{n}$ is the fraction of WR stars brighter than a
particular Type Ibc SN progenitor limit. Using all the limits in Table
\ref{obsprogenitors}, we find a probability of only 15 per cent that
we would not have seen a Type Ibc progenitor thus far. 

The WR stars in the LMC which are plotted in Figure
\ref{WRprogenitorcolours1} are a mix of massive binaries and single
stars, while it is of course a single star that explodes. However this
does not matter for our test, as if the typical WR population of the
LMC were giving rise to Type Ibc SNe we should see the progenitor
stars or their binary systems.

We have performed a similar probability calculation for the sample of
WR stars described by \cite{2012MNRAS.420.3091B,2010MNRAS.405.2737B}.
Here their sample is biased towards brighter magnitudes with no stars
in their sample below a $M_V \simeq -4$. If we assume this is
representative of the progenitor population then the probabilities in
Table \ref{stats} decrease. For example, our most restrictive event SN
2002ap would have a probability of 0.01. The probability for the
entire population would be $4\times 10^{-6}$ and so essentially zero,
even without this SN. This demonstrates how difficult it is to detect
Wolf-Rayet stars in other galaxies at great distances let alone
progenitors that may also be fainter than these Wolf-Rayet stars. It
also demonstrates that this bright end of the WR luminosity function
is almost certainly not producing the Ibc SNe that we see in the local
Universe.

\subsection{Limits on progenitors from population modelling}

\begin{figure*}
\includegraphics[angle=0, width=168mm]{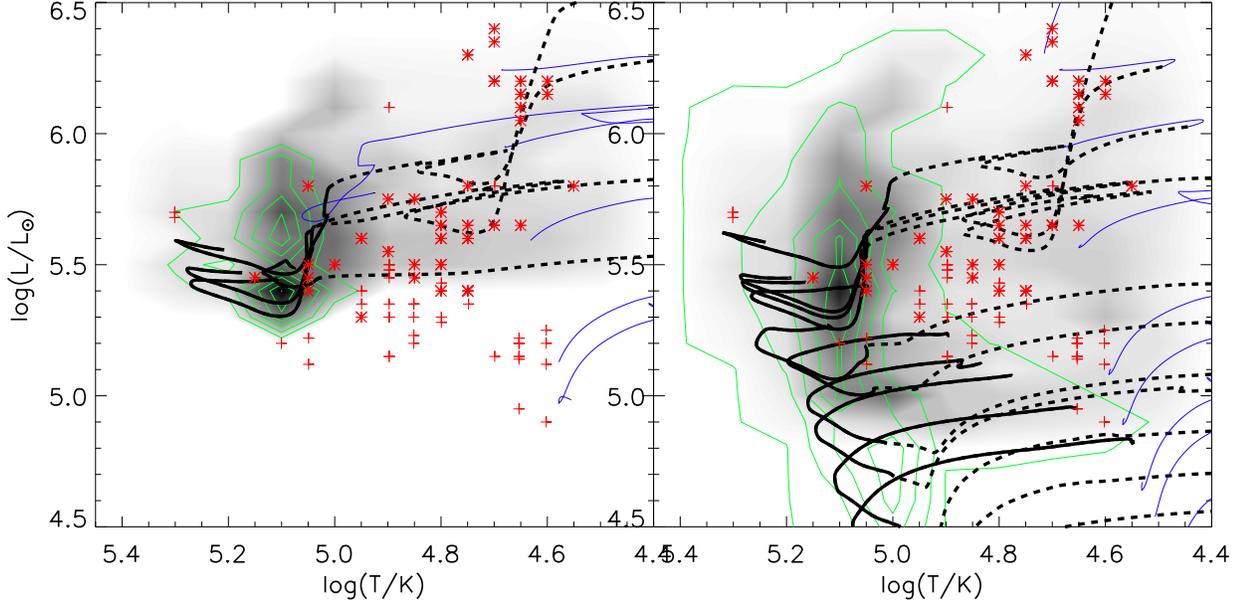}
\caption{{Theoretical HR diagrams that compare models and observations
  of WR stars. The shaded regions show the density of the population
  we expect from a synthetic population of the WR stars calculated by
  BPASS assuming constant star-formation, a Salpeter IMF and requiring
  that $\log (L_{\rm WR})>4.9$. Helium stars less luminous than this
  still contribute to SNe but are assumed not to be WR stars. The
  lines show the evolutionary model tracks, blue lines represent the
  pre-WR evolution, the dashed black lines represent when $0.01<X_{\rm
    surface}<0.4$ and the solid black lines represent when $X_{\rm
    surface}<0.01$. The left panel shows the single-star models with
  masses 27, 30, 50, 80 and 120$M_{\odot}$ . The right panel shows a
  few example binary models, the masses included are 10, 13, 15, 18,
  20, 25, 30, 50, 80 and 120$M_{\odot}$. In both panels the red points
  are observed WR stars, with WN stars indicated by `*' and WC stars
  with `+'. The data is taken from \citet{Har06} and
  \citet{San12}. The binary models cover a wider region in luminosity
  and temperature than the single-star models. The green contours
  indicate the region of the HR diagram covered by our Ibc progenitor
  populations.}}
\label{HRofwrstars}
\end{figure*}

Comparing the progenitor limits from archival imaging to the observed
population of Wolf-Rayet stars is the simplest approach to
constraining the progenitor population of Type Ibc SNe. However, it is
limited by the fact that observed Wolf-Rayet stars will lie within a
range of different evolutionary states, with some stars closer to
core-collapse than others. It is possible that Type Ibc progenitors
become fainter in optical bands as they evolve closer to the point of
core-collapse \citep[e.g.][]{Yoo12}. To account for this we compare
our observed progenitor limits to synthetic progenitor populations
created with \textsc{BPASS}. 

There is great uncertainty in creating models of WR stars, both in
atmosphere and stellar models. A great amount of work has been
performed recently improving the models of WR atmospheres
\citep[e.g][]{San12} however uncertainties still remain. Numerical
models of WR stars have also improved however there are still
{ambiguities}, mainly concerning the mass-loss rates of these
stars. Comparing our models to those of \citet{Yoo12} and
\citet{eks12} we find the evolution is qualitatively similar. There is
different evolution however towards the endpoints of the models. In
\citet{eks12} most WR models increase in effective temperature up
until core-collapse and so become optically fainter. The difference is
most obvious for more massive WR stars. This difference is likely to
be due to \citet{eks12} adopting a lower Solar metallicity and
different mass-loss rates. Those of \citet{Yoo12} and our models both
have effective temperatures that decrease slightly towards the end of
evolution. But the models of \citet{Yoo12} also reach higher surface
temperatures than our models. We assume this to be the result of the
models of \citet{Yoo12} starting on the helium main-sequence while our
models arise naturally due to binary evolution so typically experience
less mass loss. The difference could also be due to how the different
stellar evolution codes deal with the WR inflation effect as discussed
by \citet{2006A&A...450..219P} and \citet{2012A&A...538A..40G}. This
is perhaps the key question that must be resolved before we are able
to accurately predict the pre-SN structure and parameters for WR
stars. We do not use the final end-points of our models to estimate
progenitor luminosities because of the question of whether or not
progenitors are inflated. Inflation refers to the increase in the
radii of stellar models due to radiation pressure on the iron-bump in
the stellar opacity \citep{1999PASJ...51..417I}. The occurrence has
been shown to depend on metallicity and mass-loss rates used in the
stellar models \citep{2006A&A...450..219P}. The difference in radii
caused by this effect will also affect the effective temperatures of
the models. {Our method includes a range of radii for each
  model in the population. Therefore allowing, in an approximate way,
  for the uncertainty in a WR stars radii.}

It is difficult to determine how WR stars do evolve towards the end of
their lives and therefore we take this into consideration in our
analysis below. We stress while there are {many} uncertainties
in predicting the WR population our BPASS synthetic populations have
been tested against a number observations where the modelling of WR
stars is key to reproducing the observations from nearby to
high-redshift \citep[e.g.][]{EIT, ES09,E09, ES12}. To provide further
validity to our WR models in Figure \ref{HRofwrstars} we compare the
locations predicted for our single star and binary population in the
Hertzsprung-Russell diagram to those inferred from the Potsdam models
for observed WR stars in \citet{Har06} and \citet{San12}. The Figure
shows that in general our models when hydrogen-free, do match the
location of observed WR stars. {We note that the observed
  sample has complex selection effects and does not represent a
  constant star-formation history or fully sampled IMF. Therefore our
  model population can only be compared to the region of the WR stars
  on the HR diagram, not their relative distribution. We have also
  assumed to be a WR star a model must have $\log
  (L/L_{\odot})>4.9$. Stars with no surface hydrogen below this limit
  are still helium stars but are unlikely to be observed as WR stars.
  In summary, no stellar models of WR stars are perfect predictors of
  the observed WR star population but our models are at least close to
  reproducing the observational data here and in previous studies
  using BPASS.}

We create the populations using the single-star and binary star
populations from BPASS. We predict the magnitudes by matching our
stellar models to the Potsdam stellar atmosphere models of WR stars. We
note that while we used these models above in Section 5.1 our method
of using them here is separate. We match the stellar model radii and
surface temperatures to the Potsdam WR models as described on the
above website. For full details of the construction of the stellar
evolution models we refer the reader to \citet{EIT} and for the
matching of models to stellar atmospheres to \citet{ES09}.

We create four synthetic populations to compare to the progenitor
non-detection limits. For both a single star population and our
fiducial binary and single star mix we create two luminosity
distributions with the following parameters:

\begin{enumerate}
\item A WR star population with $X_{\rm surface}<0.001$ and $\log(
  T_{\rm eff}/{\rm K})>4$. This is equivalent to the observed WR
  population.
\item A possible Ibc progenitors population with $X_{\rm
  surface}<0.001$, $M>2M_{\odot}$ and {we require the
  model to have completed core helium burning. This restricts the
  population to models closer to core-collapse. This includes a range
  of radii for each model that is an approximate method to account for
  the uncertainty in the evolution of the progenitors. The population
  provides a range of possibly magnitudes for the progenitors similar
  to the range allowed from various models of WR stars.}
\end{enumerate}

We have also created a third population based on the end points of our
stellar models. {We find however that the probabilities for
  these populations reproducing the non-detections are low and similar
  to the model WR populations with values of 4.6 and 1.5 per cent for
  single stars and binaries respectively. The latter limit is low
  because of the B-band limit for SN 2002ap.  This is because our
  models become cooler and therefore slightly brighter in the optical
  towards the end of their lives. However as noted above this
  behaviour is different to that found by \citet{eks12} and
  \citet{Yoo12}.  However the probabilities are sensitive to any extra
  dust absorption that has not have accounted for in calculating the
  B-band limit. We note that if we were to consider only R-band
  limits, that are more weakly affected by dust, the probabilities
  rise to 19 and 17 per cent respectively. This difference is because
  using the endpoints along essentially describes only a line of
  points in the HR diagram rather than allowing for a range of
  values. One option would be to blur the stellar endpoints by
  considering different amount of intrinsic dust absorption or
  estimating what the radii of the WR stars if the envelope inflation
  did not occur. Attempting this would include arbitrary physical
  mechanisms that would allow us to achieve any fit we like. Therefore
  we decided to use a similar method as we employed in Paper I. Using
  population (ii) outlined above. Using the natural range of the
  temperatures and luminosities in the models themselves as they
  evolve towards the endpoint of their lives. Some vary little, such
  as the more massive WR stars, some vary more, especially the low
  mass binary stars. Hence our preferred quasi- progenitor population
  in an approximate method accounts for uncertainties in the WR models
  as well as other factors such as dust that might be produced in WR
  binaries \citep{Crow03}.}

\begin{figure}
\includegraphics[angle=0, width=0.47\textwidth]{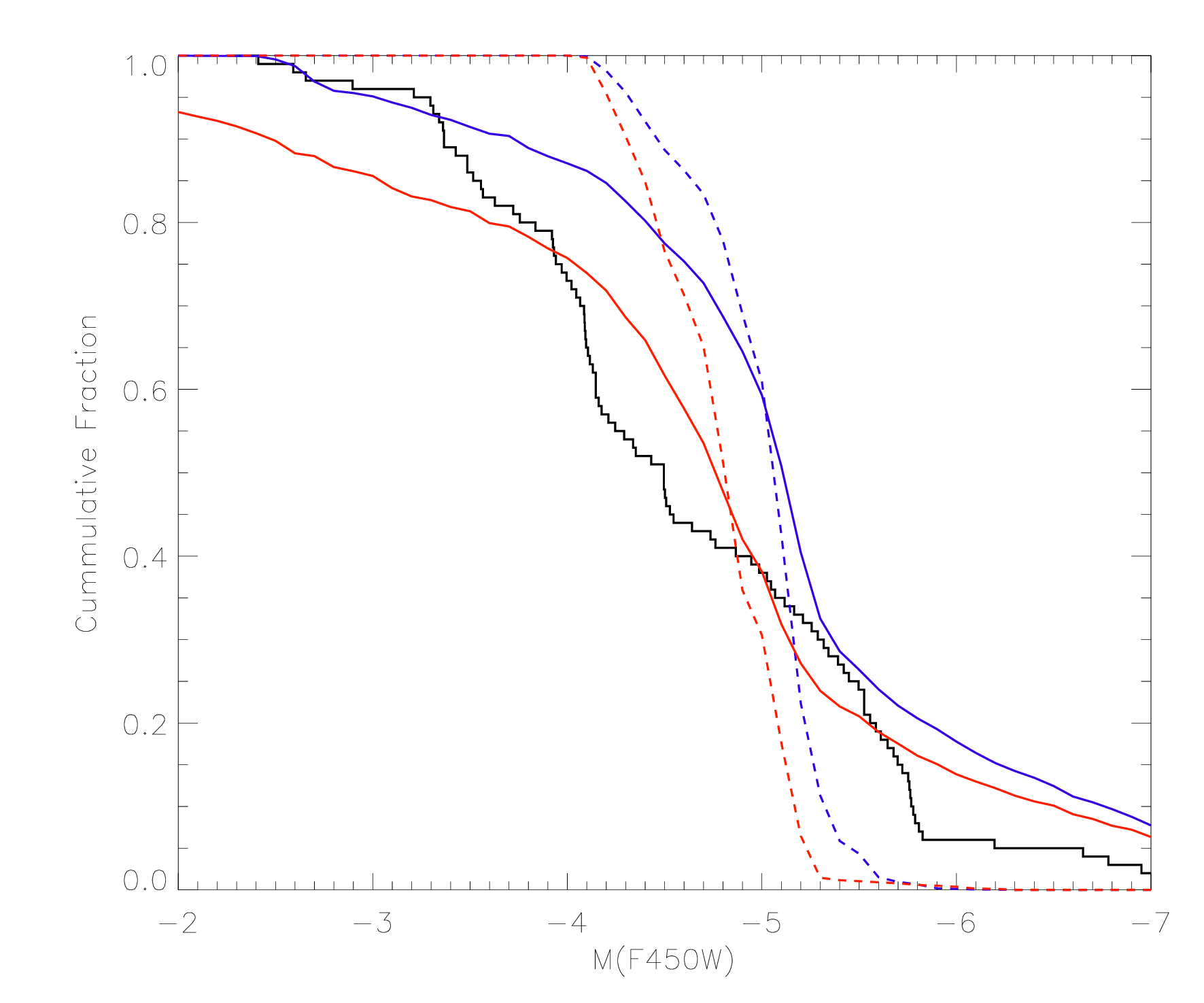}
\includegraphics[angle=0, width=0.47\textwidth]{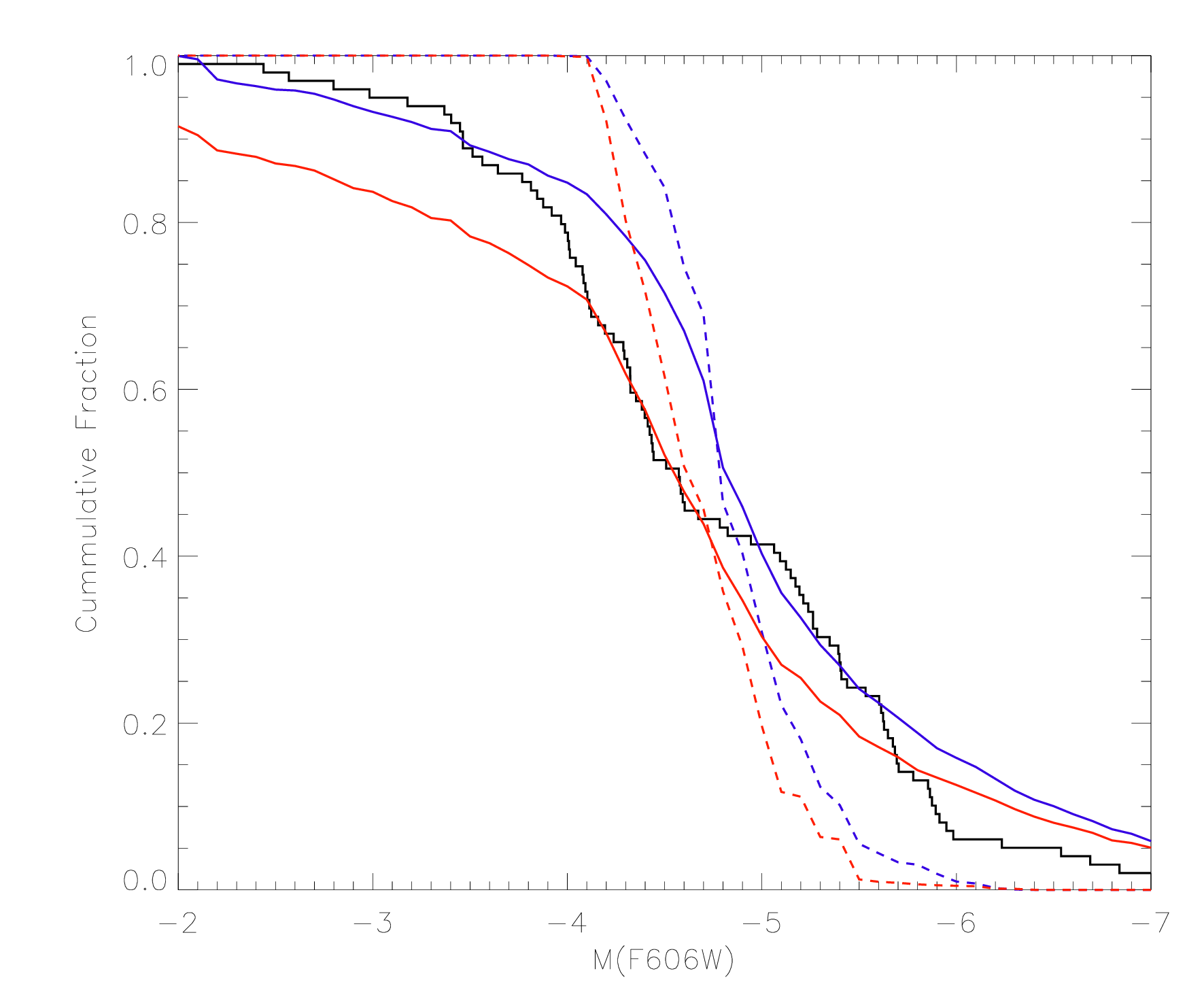}
\caption{Cumulative frequency $F450W$ (upper panel) and $F606W$ (lower
  panel) magnitude distributions for observed WR stars (black line),
  single star synthetic populations (dashed lines) and fiducial
  binary-single star mix (solid lines). The blue lines are for the
  synthetic WR population and the red lines the synthetic progenitor
  population.}
\label{imf}
\end{figure}

Comparing the different luminosity distributions predicted from the
various populations for different filters in Figure \ref{imf} we see
that for the F606W filter the observed and synthetic WR populations
are similar. {We note the observed population here is the same
  as we use in Section 5.1 and not those in Figure 9. The match for
  the F450W filter is slightly worse.} We suggest this is due to
intrinsic dust extinction in the observed WR stars that is greater
than that of the foreground extinction. We find a better match between
the models and observations if we were to include on average, $A_{\rm
  V}=0.5$, for the synthetic population but we do not apply it
here. {Furthermore we see that our predicted progenitor
  magnitude range for the binary populations are greater than those
  for the single star WR populations population. This is because there
  is a wider range of helium and WR star masses allowed from binary
  evolution. Then the synthetic populations for the progenitor models
  are fainter than the synthetic WR populations. This is because the
  progenitor stars are hotter and less luminous than the general WR
  population. After core-helium burning a WR star's luminosity is at the
  lowest point during its entire evolution.}

The SN types resulting from the stars in these synthetic populations
are determined using the parameters listed in Table \ref{snpram}. For
the binary population we include the contribution of the primary and
secondary stars to the WR population. We also consider the companion
flux when calculating the observed magnitude of the system. We see in
Figure \ref{imf} that the synthetic single-star WR population has a
similar mean magnitude to the observed LMC WR population used above,
but is not able to reproduce the spread of magnitudes. In contrast,
the synthetic binary WR population is a much closer match to the
observed distribution. This is because binary interactions increase
the amount of mass-loss a star can undergo, and so widens the range of
initial masses (and hence luminosities) which give rise to WR
stars. In addition the inclusion of the flux from the secondary in
binary systems leads to some WR stars being apparently more luminous
than possible from single stars alone.

We again stress that our method differs from that of \citet{Yoo12},
who only consider the very end point of each evolutionary track as the
point of explosion. We use the predicted WR population and the
population after core-helium burning. {In using these we are
  allowing for our synthetic progenitors to cover a region of the HR
  diagram rather than a narrow line of the end points. The advantage
  of this is that it gives us a range of possible magnitudes for each
  model, and hence allows us to take account of the uncertainties of
  stellar evolution and Wolf-Rayet atmosphere modelling in our
  analysis in an approximate method rather than calculating numerous
  sets of binary models with varying input physics and mass-loss
  rates. Such a calculation would require 100,000's of stellar models
  and is beyond the scope of this paper.} The method of \citet{Yoo12},
for example, is not incorrect in that they report the results of the
end points of the calculations, but one has then to assume that these
models are perfectly valid and describe accurately the unobserved
processes in the last few thousand years before
core-collapse. {Hence we suggest that our approach approximates
  modelling uncertainties and we assume that we cannot definitively
  predict the end points. Especially considering we are attempting to
  model a population of non-detections.}

We also note that our predicted magnitudes are different to those of
\citet{Yoo12}. This is because that work is based on the narrow band,
$v$, rather than the broad-band filter, $V$, commonly available for
progenitor detections. The narrow band filter does not include
emission line flux which can contribute significantly to the
luminosity of Wolf-Rayet stars, and $v$ can be typically 0.5-1.0
magnitudes fainter than broadband $BVR$. This was also pointed out by
\citet{Yoo12} who convert between the two filters by assuming $V=
v-0.75$. In our method by calculating $V$ directly we account for the
variation in how the two magnitudes are related depending on effective
temperature of the WR stars.

Using these synthetic populations we can again compare the progenitor
limits to the various synthetic populations. From each SN we can
calculate a probability that the progenitor would remain undetected
from the fraction of the synthetic population with a magnitude less
than the limit. We then combine those individual probabilities to
estimate the probability that no progenitor would have been observed
in any of the pre-explosion images. We show the resulting
probabilities and their combination in Table \ref{stats}. As mentioned
previously, the probability that we would not have detected a Type Ibc
progenitor if the LMC WR sample is representative of the progenitors
is 0.16. Similarly low values are found for all the progenitor
models. The lowest values are for our WR populations and the
progenitor population. The highest probabilities are for our
single-star and binary populations from the end of core-helium burning
to the end of the model, with 0.13 and 0.12 respectively. 

All these low values are the result of the B-band limit for SN
2002ap. If we however account for the possibility of intrinsic dust
around WR stars or switch the the R-band limit which is less affected
by extinction these probabilities would increase. We note that for the
binary populations the probability of detecting SN 2002ap is nearly
double that for the single-star populations. The reason why there is
little difference between the overall probabilities is that the binary
stars also predict more luminous progenitors than the single-star
population. This perhaps gives the false impression that single stars
are more likely to be the possible progenitors. If we consider only
the deepest limits SNe 2002ap, 2004gn and 2010br then the binary
population again are slightly more favoured.

In conclusion, while we favour a mixed population of Wolf-Rayet stars
and lower mass helium stars in binaries as the progenitors of Type Ibc
SNe, we cannot set strong constraints from the current limits and
non-detections. We are also limited by our limited understanding of WR
star evolution and which stellar models correctly reproduce the
observed population of WR stars and progenitors of type Ibc SNe.

{One thing that is apparent is the low probabilities we obtain
  indicate that the Ibc progenitor population is fainter than expected
  from the magnitudes of WR stars. Therefore other factors could be at
  play making it more difficult to observe Ibc progenitors.} For
example, either dust being produced to decrease their luminosity
\citep[e.g.][]{Crow03} or that the predictions of \citet{Yoo12,eks12}
are correct and the progenitors should be hotter at the time of
core-collapse than we predict.

\begin{table*}
\caption{The probabilities that no progenitor would have been observed given the different models.}
\label{stats}
\begin{tabular}{@{}lccccccccccccc}
\hline
\hline
		&   		&       		&		\multicolumn{4}{c}{Progenitor models} 			\\
		& SN   	&   Observed 	& {Single}& {Single}& 	{Binary}&	{Binary}\\
SN  		& Type 	&  WRs  	& WR stars  & Post-He burn   &WR stars  & Post-He burn    \\
\hline
2000ew 	& Ic  		&    0.96   &   1.00   &   1.00      &       0.93  &   0.94     \\
2001B  	& Ib  		&    1.00   &   1.00   &   1.00      &       1.00  &   1.00     \\
2003jg 	& Ibc	        &    0.94   &   1.00   &   1.00      &       0.86  &   0.89     \\
2004gt 	& Ibc  		&    0.94   &   0.98   &   0.99      &       0.83  &   0.87     \\
2005V  	& Ibc	        &    0.99   &   1.00   &   1.00      &       1.00  &   1.00     \\
2007gr 	& Ic            &    0.97   &   1.00   &   1.00      &       0.90  &   0.92     \\
2011am 	& Ib  		&    0.99   &   1.00   &   1.00      &       0.97  &   0.97     \\
2011hp 	& Ib  		&    0.99   &   1.00   &   1.00      &       0.97  &   0.97     \\
2012au   & Ib           &    0.99   &   1.00   &   1.00      &       0.97  &   0.98     \\
\hline
2002ap 	& Ic  		&    0.48   &   0.089  &   0.18      &       0.21  &   0.35     \\
2004gn 	& Ic  		&    0.76   &   0.95   &   0.99      &       0.77  &   0.82     \\
2010br 	& Ibc	        &    0.57   &   0.62   &   0.73      &       0.55  &   0.66     \\
\hline
  \multicolumn{2}{l}{Model Probability}  &    0.16  &   0.051  &   0.13     &      0.048 &   0.12   \\
 \multicolumn{2}{l}{02ap, 04gn, 10br only} &  0.21 & 0.052 & 0.13 &  0.089 & 0.19 \\

\hline
\hline
\end{tabular}
\end{table*}

\section{Discussion}

There are two main candidates for the progenitors of Type Ibc SNe:
Wolf-Rayet stars (M$\ga$20 \msun) and lower mass helium-stars. The
former will experience mass loss which is primarily due to stellar
winds, and will effectively evolve as single stars. The latter are the
result of close binary evolution, where the mass-loss is due to binary
interactions of Roche-lobe overflow or common envelope evolution. Both
progenitor channels are plausible.  WR stars are commonly observed in
star-forming environments, but to the best of our knowledge no
completely hydrogen-free low-mass helium stars are known (with a core
mass that will create an ONeMg or Fe-core and experience
core-collapse) and their existence remains conjecture.

\subsection{Arguments in favour of interacting binaries as progenitors
  of Ibc SNe}

From the limits presented in Section \ref{s5}, it is difficult to
distinguish between the different progenitor populations. However
considering the deepest limit alone, SN 2002ap, it is more likely to
have been detected if Wolf-Rayet stars were the sole progenitors of
Type Ibc SNe. This one progenitor was less likely to have been
observed if it was a binary system as stated by \citet{Cro08}. The
number of non-detections to date suggest that we do not fully
understand the evolution of hydrogen-free stars. We do suggest however
because of the lower magnitudes of WR stars as they approach
core-collapse as also discussed by other authors that at the moment
the number of non-detections are not enough to constrain the
population directly.

The uncertainties in the models which make a comparison difficult
include how the stars evolve after core-carbon burning, and how much
dust might a low-mass helium star create in a binary.  Systems such as
those discussed by \citet{Crow03} would be difficult to observe, and
alter our predicted luminosity distributions significantly. Therefore
from the progenitor detections we can conclude there is only
{very} weak evidence from the deepest non-detections that
binaries are the more favoured progenitors.

The strongest case for binary stars contributing to the type Ib/c SN
progenitor populations can be made by the fact we can reproduce the
observed relative rate of different SN types
\citep{Pod92,1998A&A...333..557D,Yoo11}.  Current single star models predict too
many Type II SNe, however recent studies have indicated that the
mass-loss rates of red supergiants remain a large source of
uncertainty \citep{yc10,georgy}.

Currently no single star model can fit the population of RSGs, WR
stars and SNe rates simultaneously. The Geneva rotating stellar
evolution models have come very close \citep{eks12}, but assume a
single initial rotational velocity rather than a distribution of
rotation velocities as are observed \citep{Hunt08}. In contrast,
binary models can also come close to reproducing both observed stellar
populations and relative SN rates with reasonable distribution of
initial binary parameters
\citep[e.g.]{1998NewA....3..443V,2003A&A...400..429B,EIT,ES09}. Also
they are able to match individual binary systems that have been
observed \citep[e.g]{E09}. In our binary population model which fits
best both the observed relative SN rates and the non-detection of a
Type Ibc progenitor, we find that single WR stars contribute one-fifth
of Ibc SNe, with binary systems giving rise to the rest. Furthermore
in our synthetic population only half of the SNe will be in a binary
at the time of core-collapse, the remainder being apparently
single. This is because binaries are either unbound in the first SN or
have a compact companion at the time of the second SN. Deep
post-explosion images with the {\it Hubble Space Telescope} at the
sites of Type Ibc SNe may help identify the surviving binary companion
of the progenitor as described by \citet{2009ApJ...707.1578K}.

It is important to note that our binary population model also predicts
the location on the HR diagram of the progenitors of hydrogen-poor
Type IIb and IIL SNe. In Figure \ref{HR} we show a cartoon HR diagram
based on our synthetic population, where the YSG progenitors lie above
our predicted location for the hydrogen-deficient SNe. We note that
this matches the observed progenitors of SNe 1993J, 2008ax, 2009kr and
2011dh, which are intermediate between the Type Ibc progenitors and
hydrogen rich IIP RSG progenitors.

\begin{figure}
\includegraphics[angle=0, width=84mm]{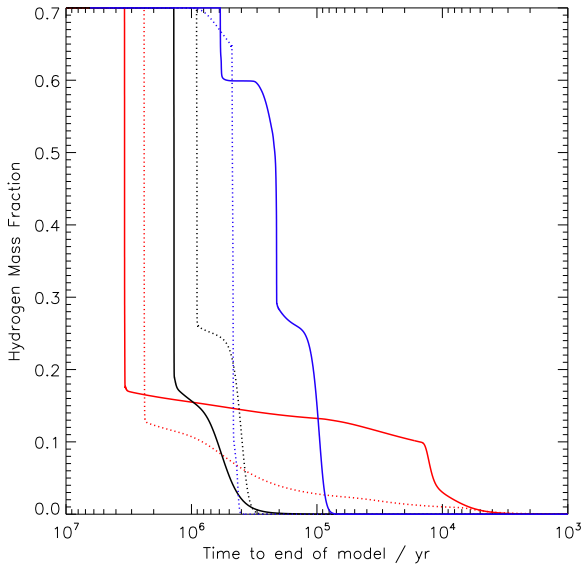}
\caption{Diagram of the hydrogen mass fraction of different progenitor
  models before the end of our models. The solid and dotted blue lines
  are for 30\msun\ and 50\msun\ single-star models respectively. The
  solid and dotted black lines are for 15\msun\ and 20\msun\ binary
  models respectively. The solid and dotted red lines are for
  10\msun\ and 12\msun\ binary models respectively. .}
\label{endtimeplot}
\end{figure}

\begin{figure*}
\includegraphics[angle=0, width=168mm]{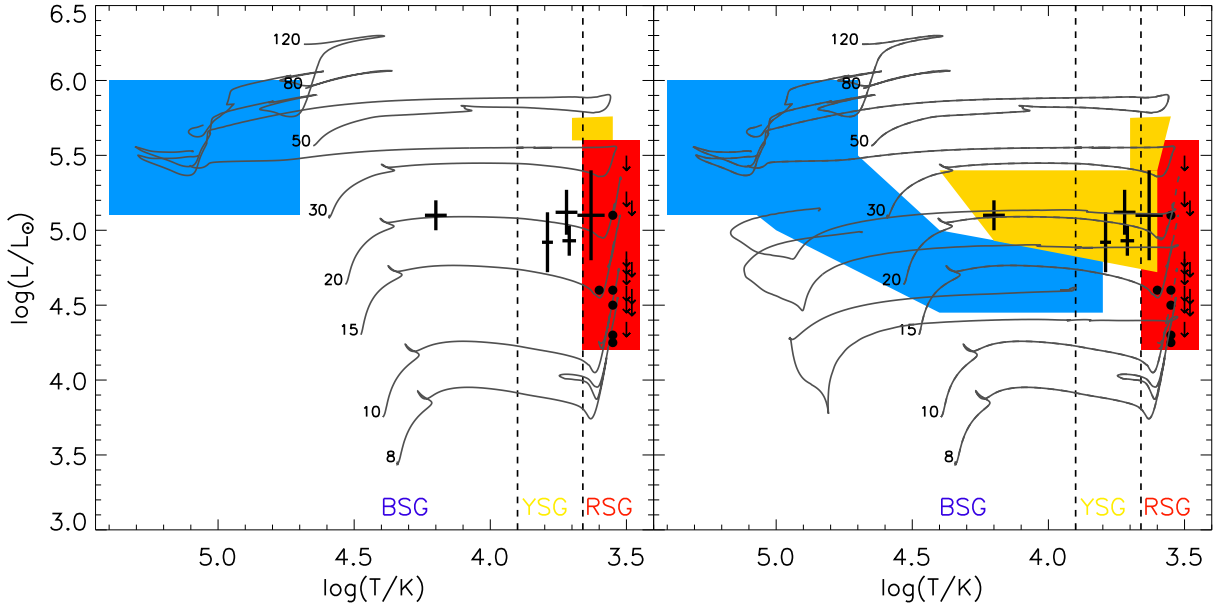}
\caption{Cartoon HR diagrams of SN progenitors, the red, yellow and
  blue regions show the expected location of progenitors for type IIP,
  other type II and type Ib/c SNe. The left panel shows the
  single-star scenario. The solid lines show the evolution tracks for
  stars with masses given at their initial location.  The right panel
  shows the binary scenario with the solid tracks at 10, 15 and
  20\msun\ showing binary evolution tracks and the dashed lines the
  single star tracks. In both plots the points with error bars show
  the locations of SNe 1987A, 1993J, 2008cn, 2009kr and 2011dh
  \citep{1992PASP..104..717P,2004Natur.427..129M,2009ApJ...706.1174E,2010ApJ...714L.254E,Fra10b,Mau11}
  respectively. The circles show the progenitor locations of observed
  type IIP progenitors and the arrows the upper limits for these
  progenitors \citep{Sma09b}.}
\label{HR}
\end{figure*}

Hence the question then is, where are the binary systems that harbour
Ibc progenitor systems?  If we examine our binary evolution models
more closely we find that stars below approximately 15\msun\ spend
only a very small fraction of their total lifetime without hydrogen in
their structure before core collapse. This is shown in Figure
\ref{endtimeplot} which displays the hydrogen mass fraction of
single-star and binary models before the end of the stellar
models. Stars less massive than 15\msun\ retain a low mass hydrogen
envelope that is helium rich. This is only lost a short time before
core collapse. In the right-hand panel of Figure \ref{HR} the envelope
is lost as the stars evolve from their hottest extent to cooler
effective temperatures. Stars with parameters similar to these stars
have been observed and are objects such as V Sagittae, WR7a and
HD45166\footnote{HD45166 is slightly more hydrogen rich than predicted
  by our models. This is probably a result of the secondary star
  filling its Roche lobe at periastron and transfering hydrogen back
  onto the primary \citep{hd45166a}.}
\citep{vsag,wr7a,hd45166a,hd45166b}.  The objects are often discarded
as potential Ibc progenitors due to the presence of hydrogen on their
surface.  In Figure \ref{endtimeplot}, however we show the surface
hydrogen mass fraction for our models which indicate that these
possible type Ibc progenitors are hiding large helium cores and the
residual hydrogen could be stripped in the last $10^{4}$\,yrs of their
life; hence it may not be surprising that they exist with in this
configuration and that completely hydrogen free He-stars in binaries
are difficult to locate. This may seem unexpected but we stress that a
binary interaction does not immediately lead to complete loss of the
hydrogen envelope in a star. Roche-Lobe overflow or common envelope
evolution removes most of the hydrogen envelope but eventually an
envelope mass is reached at which point it collapses back within the
star's Roche Lobe.  Stellar winds are then responsible for removing
the remaining gas. For stars above 15\msun\ the winds are strong
enough to do this during helium burning. Below this limit it is not
until the star evolves to a helium giant that the mass-loss rate
becomes strong enough to drive off the hydrogen before
core-collapse. We note that at lower metallicity weaker winds may
reduce the number of type Ibc SNe from these binary stars, therefore
the type Ibc rate will decrease at lower metallicities.

The other piece of observational evidence that supports helium stars
as SN progenitors is the ejecta masses determined from the studies of
the lightcurves and observed ejecta velocities. There are now many Ibc
SNe with well sampled lightcurves, and some with data stretching
through the nebular phase. Simple analytical models following
\cite{1982ApJ...253..785A} can estimate ejecta mass and $^{56}$Ni mass
from the rise time, width and peak of the lightcurve. More
sophisticated radiative transfer modelling with Monte Carlo techniques
\citep[for example][]{2001ApJ...550..991N,Maz07,dessart} have been
used to determine also the kinetic energy from the velocity
information in the spectra. A summary of the main results for type Ibc
SNe is compiled in Table\,\ref{tab-LCs}.  The models of the most
common Ibc SNe types in the local Universe (within our 28\,Mpc
volume), and those for which we have constrained the properties of the
progenitors, all tend to be between 1-3\msun.  This fact has recently
been highlighted by sophisticated modelling of Ibc SNe by
\cite{2012MNRAS.422...70H} and \cite{2012MNRAS.424.2139D}. In
addition, \cite{drout} presented a homogeneous study of the ejecta
masses of twenty-five Ibc SNe and found similarly low masses
consistently through the sample.  There are a small percentage of the
overall Ic SNe class that have very broad lightcurves and usually, but
not always, high ejecta velocities (often called Ic-BL, or broad-lined
Ic). There is some evidence that they have significantly higher ejecta
masses. Their lightcurves are significantly broader and a higher
ejecta mass is the obvious way to explain the differences with a
similar physical model.  However these Ic-BL are rare, with only one
(SN 2002ap) having been discovered within 28\,Mpc in the duration of
our survey.  Their contribution to the overall Ibc SN population is
likely less than 3 per cent (see Section 2) and these could
conceivably be from higher mass He or CO stars. Wolf-Rayet stars would
be the obvious candidates for some of these type Ic SNe
\citep{2012MNRAS.424.2139D}.

In Table\,\ref{tab-LCs} we also list the oxygen ejecta mass where it
has been estimated by model application. Generally this is around
1\msun\ or lower, also supporting the idea that the bulk of the Ibc SNe
are not from WR stars progenitors with zero-age main-sequence masses
$\ge25$\msun\ \citep[see][]{Maz10}.

\begin{table*}
\caption{Measured ejecta masses for Ibc SNe from the literature. These
are lightcurve modelling. In some cases an estimated mass of oxygen is
also determined from spectral modelling and is added here for
reference. Typical stellar evolution models predict that a 25\msun\
star would produce between 2-4\msun\ of oxygen during its
nucleosynthetic evolution.}
\label{tab-LCs}
\begin{tabular}{lcccl}\hline\hline
SN        &  Type   &  O mass / \msun\   & Total ejecta mass / \msun\      &   Refs\\\hline
2007Y   &   Ib      & 0.2         & 0.4                             & \citet{2009ApJ...696..713S}\\
1994I    &   Ic      & ..             & 0.7                            & \citet{2008MNRAS.383.1485V} \\
2007gr &   Ic       &  0.8         &  1-2                         & \citet{Maz10}, \citet{Hun09}\\
2006aj  &  Ic       & 1.3           & 2                              & \citet{2006Natur.442.1018M}, \citet{2007ApJ...658L...5M}\\ 
2002ap &  Ic       &  1.2          & 2.5                           & \citet{Maz07} \\
2003jd  &  Ic       & ...            & 3                              & \citet{2008MNRAS.383.1485V} \\
2008D   & Ib        & 1.1         & 3.6                            & \citet{2012MNRAS.424.2139D} \\
2004aw  &  Ic       & ...            & 5                              &  \citet{2008MNRAS.383.1485V} \\
2009jf &  Ib  &    ...              & 5-7                           & \citet{2011MNRAS.416.3138V}\\
1997ef    & Ic      &   ...           & 7-10                       & \citet{2000ApJ...545..407M}\\
1998bw & Ic       & 2-3              & 10                         & \citet{2001ApJ...550..991N} \\
2011bm & Ic      & 5-10         &   7-17                      & \citet{2012ApJ...749L..28V}\\
\hline 
Sample Average & Ib      & ...& 2$^{+1.1}_{-0.8}$ &   \citet{drout} \\
Sample Average & Ic      & ...& 1.7$^{+1.4}_{-0.9}$ &   \citet{drout} \\
Sample Average & Ic-BL & ... & 4.7$^{+2.3}_{-1.8}$ &   \citet{drout} \\
\hline \hline
\end{tabular}
\end{table*}

We predict ejecta masses from the synthetic populations as described
in \citet{ETsne}. The should be considered upper limits as we do not
evolve our models to Fe-core formation and so may underestimate the
final remnant mass. Our binary progenitor models (which mainly form
neutron-star remnants) have an average ejecta mass of
$4.2\pm2.4$M$_{\odot}$ while single star progenitors (which mainly
form black-hole remnants), have ejecta masses in the range to
$6.6\pm0.9$M$_{\odot}$. This further supports the idea that
\textit{normal} type Ibc SNe come from helium stars formed in binaries
with initial masses below $20$M$_{\odot}$ and form neutron stars at
core collapse.

\subsection{Arguments in favour of Wolf-Rayet stars as progenitors of
  Ibc SNe}

Recently, much work has been done on the stellar populations near the
location of SNe \citep[e.g.][]{Mur11,leloudas,And12}. These range from
studying the resolved stellar populations using high resolution HST
data, such as that presented here \citep{Mur11} to lower resolution
ground-based data that typically use H$\alpha$ to probe regions of
star formation.  Using the latter technique \cite{And12} propose that
type Ibc SNe tend to show stronger spatial association with H$\alpha$
emission in galaxies compared to type II SNe, and type II-P in
particular.  As H$\alpha$ emission originates in young clusters and
associations, their interpretation is that type Ibc SNe may arise from
significantly higher mass stars than type II-P. They also find that
the spatial correlation is strongest for type Ic SNe, even more so
than type Ib. \cite{And12} interpret this as an increasing mass
sequence, with type II SNe coming from the lowest mass stars and Ib
and Ic from progressively higher mass.  This would argue in favour of
Ibc progenitors being significantly higher mass stars than progenitors
of II-P SNe, and since we know that II-P SNe arise in 8-17\msun\ stars
\citep{Sma09b,2009ARA&A..47...63S} then this might suggest WR
progenitors of type Ibc SNe with initial mass $\geq$20\msun.

These SNe are typically at distances of 50-100\,Mpc and the
ground-based imaging does not specifically identify the host H{\sc ii}
region, a fact which led \cite{2012MNRAS.tmp..183C} to study the
environments of the closest 39 CCSNe in nearby ($\leq$15\,Mpc) and low
inclination galaxies. While Ibc SNe are nearly twice as likely to be
associated with nearby H\,{\sc ii} emission,
\cite{2012MNRAS.tmp..183C} finds that these nebulae are long lived,
giant H\,{\sc ii} regions with typical lifetimes more than
20\,Myrs. Hence the association only provides weak constraints on Ibc
progenitor masses.  Binary evolution would extend the period during
which Type Ibc SNe occur up to 20 Myrs and this is still roughly in
agreement with observed population ages from the H\,{\sc ii} region
analysis.  The greater mean age of binary SN progenitors is also in
accord with the absence of any H{\sc ii} regions or large young
stellar populations at the site of the Type Ibc SNe in our sample. The
typical lifetime of a H{\sc ii} region is between 3 to 5 Myrs,
i.e. shorter than the lifetime of most binary Type Ibc
progenitors. Only the most massive regions such as NGC604 or 30
Doradus have more extended periods of star-formation, and such systems
are rare.


There are two other factors that may have prevented us from detecting
WR progenitor stars.  Firstly it is possible that WR stars evolve
rapidly in the latter stages of their lives.  This could mean that,
just prior to core-collapse, they appear significantly different to
the WR stars we observe in the LMC, possibly getting hotter and
fainter in the last 5 per cent of their lifetimes.  This has been
investigated by \citet{Yoo12} and \citet{eks12} indicates similar
evolution. However we suggest that progenitors are slightly more
observable than this. One reason for the difference is due to
\citet{Yoo12} using mainly magnitudes in the narrow band $M_{\rm v}$
with a constant correction factor to $M_{\rm V}$ rather than allowing
for the range of differences between the magnitudes of 0.5-1.0. Also
their models end their lives with higher surface temperatures than we
find which again makes the model progenitors dimmer at optical
magnitudes. Secondly, we may have underestimated the extinctions
towards the WR star progenitors (and hence our detection limits are
not as deep as we have presented here), or WR stars could create and
eject dust in the last fraction of their lifetimes. The values for
$E(B-V)$ used in Table\,\ref{obsprogenitors} are those estimated for
the foreground line of sight (including host galaxy extinction where
possible).  The distribution is characterised by a mean and standard
deviation of $<E(B-V)>=0.22\pm0.28$, which is similar to the
extinctions estimated by
\cite{2012MNRAS.420.3091B,2010MNRAS.405.2737B} towards WR stars in
NGC5068 and NGC7793. In addition, we have added our estimates of
extinction to the limiting magnitudes, whereas for the LMC sample we
have simply taken foreground Milky Way extinction. Any extra internal
LMC, or circumstellar extinction would tend to make the LMC reference
sample brighter for our comparison purposes and the restrictions
tighter.  In this study we have no ability to disentangle the
foreground extinction from circumstellar, hence we have not employed
the new and physically consistent methods of
\cite{2012ApJ...759...20K}. If a Ibc progenitor were found then it
would be necessary to consider the circumstellar extinction separately
to foreground ISM as discussed by \cite{2012ApJ...759...20K} as is the
case for SN2012aw.

\section{Conclusions}

We have used the observed limits on the progenitors of Type Ibc SNe in
pre-explosion images, the observed relative rates for different SN
types and a binary population model to study their progenitor
population.  From both sets of data, we find that it is difficult to
constrain the progenitor population. However for the SN with the
deepest limits, SN 2002ap, a single-stellar population is
disfavoured. Because of this and attempting to match the observed
relative SN rates we suggest that the bulk of the type Ibc SNe most
likely arise from low mass progenitors which have been stripped by a
binary companion. This scenario is also supported by the fact that the
ejecta masses from Ibc SNe are, in general, too low to be from massive
Wolf-Rayet stars.

The favoured progenitors are therefore the hypothetical low-mass
helium stars. However no such hydrogen-free stars with masses large
enough to experience core-collapse are known in the Galaxy. Our
stellar models of these systems indicate that this is actually to be
expected. The binary interaction that seals these stars to their final
fate does not remove all hydrogen from the stellar surface and leaves
approximately a few times 0.1\msun . This is removed by stellar winds
as the star evolves at the end of core helium burning. These objects
would appear similar to observed binary systems such as V Sagittae,
WR7a and HD45166. Therefore we suggest that the final fate of these
systems will be type Ibc SNe. Further detailed study and modelling of
these systems are required to confirm this.

While there are some Ic SNe which have broad lightcurves and likely
large ejecta masses, they are much rarer by volume compared to the
bulk of the normal Ibc SN population. It's still possible that these
Ic SNe do come from massive Wolf-Rayet stars, however
\cite{2004ApJ...607L..17P} show that even if one assumes that only
stars above 80\msun\ produce broad-lined and broad lightcurve Ic SNe
then the observed rate of such SNe is still too low.  Hence the
massive, classical Wolf-Rayet population does not produce the majority
of the normal Ibc SNe we see within 28\,Mpc. However they could
contribute to this population or supply progenitor systems producing
rarer type Ibc events.  It's possible that some fraction of WR stars
form black holes in such a way that they produce faint SNe or no
visible display.

\section{Acknowledgements}

The research leading to these results has received funding from the
European Research Council under the European Union's Seventh Framework
Programme (FP7/2007-2013)/ERC Grant agreement n$^{\rm o}$ [291222] (PI
: S. J. Smartt). The research of JRM is supported through a Royal
Society University Research Fellowship. We thank Erkki Kankare for
information on SN 2005at. We thank Nancy Elias-Rosa and Max Strizinger
for sharing spectra of 2011br and 2011am. We thank Brad Cenko for
information on SN 2004gn. We thank Stefano Benetti and Nancy Elias
Rosa for access to an NTT image of SN2011am and a spectrum of SN2011hp
both from the ESO Large Programme 184.D-1140 'Supernova Variety and
Nucleosynthesis Yields'. We thank Andreas Sander for clarifying some
details of the Potsdam WR atmosphere models. We also thank the
anonymous referee, Dirk Lennarz, Lev Yungelson, Christopher Kochanek,
Danny Van Beveren, Alexander Tutukov, Luca Izzo and Jorge Rueda for
constructive comments that improved this paper.

\bsp
\onecolumn
\appendix

\section{Complete list of SNe employed in the survey for progenitors}
\label{a1}
\small

\setlongtables
\begin{longtable}{cccccp{6cm}}
\caption{Core collapse SNe discovered between 1998-2012.25 in galaxies
  with recessional velocities less than 2000
  kms$^{-1}$ \label{table:fullsample}. The HST FOV column notes if the
  host galaxy has been observed by HST prior to explosion and if the
  position of the SN is ``in'' or ``out'' of the camera
  field-of-view.}\\
\hline
\hline 
Supernova & Galaxy & $V_{vir}$ & Type & HST FOV & Comments \cr\hline \endfirsthead
\multicolumn{6}{l}{Table~\ref{table:fullsample}
  continued}\\ 
\hline\hline 
Supernova & Galaxy & $V_{vir}$ & Type & HST flag & Comments \cr \hline \endhead \hline
\multicolumn{6}{r}{\small\sl continued on next page} \endfoot \hline
\endlastfoot \noalign{\smallskip} {\bf Core-collapse} \\ 

1998A &IC2627 & 1974 & IIpec & ...  & \citet{2005MNRAS.360..950P}SN1987A-like \\ 
1998S & NGC3877 & 1115 & IIn & ...  & \citet{2001MNRAS.325..907F} \\ 
1998bm & IC2458 & 1810 & II & ...  & \citet{1998IAUC.6882....1L}, unknown subtype \\
1998bv & PGC2302994 &1794 & II-P & ...  & \citet{1998IAUC.6900....1K}\\ 
1998dl & NGC1084 &1298 & II-P & ...  & \citet{1998IAUC.6994....3F}\\ 
1998dn & NGC 337A &1004 & II & ...  & \citet{2000AJ....120..367G}, unknown subtype\\ 
1999B & UGC7189 & 1962 & II & ...  & \citet{1999IAUC.7089....3F},unknown subtype\\
1999an & IC755 & 1605 & II & in & \citet{1999IAUC.7124....1W}, unknown subtype\\ 
1999bg & IC758 & 1537 & II-P & ...  &\citet{1999IAUC.7137....1A}, amateur LC \\ 
1999br & NGC4900 & 1013 & II-P & in &\citet{2004MNRAS.347...74P} Low luminosity II-P \\ 
1999el &NGC6951 & 1704 & IIn & out & \citet{2002ApJ...573..144D} \\ 
1999em &NGC1637 & 615 & II-P & out & \citet{2002PASP..114...35L}\\ 
1999eu &NGC1097 & 1066 & II-P & out & \citet{2004MNRAS.347...74P} Low luminosity II-P \\ 
1999ev & NGC4274 & 1089 & II-P & in & \citet{1999IAUC.7306....2G}, amateur LC.\\ 
1999ga & NGC2442 & 1168 & II-L & ...  & Peculiar II-L \citet{2009AA...500.1013P}\\ 
1999gi & NGC3184 & 765 & II-P & in &\citet{2002AJ....124.2490L}\\ 
1999gn & NGC4303 & 1616 & II-P & ...  & \citet{2004MNRAS.347...74P} possible low luminosity II-P \\ 
1999gq & NGC4523 & 364 & II-P & ...  & \citet{1999IAUC.7339....2J}, amateur LC \\ 
2000db & NGC3949 & 1020 & II-P & out & \citet{2000IAUC.7481....2P}, amateur LC \\ 
2000ds & NGC2768 & 1622 & Ib & in & \citet{2000IAUC.7511....2F}\\ 
2000ew & NGC3810 & 1061 & Ic & in & \citet{2002PASJ...54..905G} \\ 
2001B & IC391 & 1816 & Ib & in &\citet{2001IAUC.7577....2C}, Tsvetkov \citet{2006PZ.....26....3T}\\ 
2001X & NGC5921 & 1582 & II-P & ...  & \citet{2006PZ.....26....3T}\\ 
2001ci & NGC3079 & 1343 & Ic & in & \citet{2001IAUC.7638....1F}\\ 
2001du & NGC1365 & 1408 & II-P & in & \citet{2003MNRAS.343..735S}, \citet{2003PASP..115..448V} \\ 
2001fv & NGC3512 & 1504 & II-P & ...  & \citet{2001IAUC.7756....4M}, amateur LC \\ 
2001fz & NGC2280 & 1720 & II-P & ...  & \citet{2001IAUC.7759....2M}, amateur LC \\ 
2001gd & NGC5033 & 1081 & IIb & out & \citet{2007ApJ...671..689S} \\ 
2001ig & NGC7424 & 754 & IIb & out & \citet{2006MNRAS.369L..32R}, \citet{2007ApJ...671.1944M} \\ 
2002E & NGC4129 & 1152 & II & ...  & \citet{2002IAUC.7800....4M}, unknown subtype \\ 
2002ao & UGC9299 & 1607 & Ic & ...  & \citet{2008MNRAS.389..113P}: narrow He lines\footnote{SNe 2002ao and 2006jc have been termed Ibn as they show narrow He lines due to circumstellar He rich shells. We adopt Ic rather than Ibn as that more accurately reflects the nature of the event.} \\ 
2002ap & NGC628 & 686 & Ic & out & \citet{2002ApJ...572L..61M} \\ 
2002bu & NGC4242 & 737 & IIn & out & \citet{2002IAUC.7864....4A}, LC appears like II-L \\ 
2002hc & NGC2559 & 1370 & II-L & ...  & \citet{2002IAUC.7999....1W}, similar to 1979C \\
2002hh & NGC6946 & 318 & II-P & out &\citet{2006MNRAS.368.1169P} \\ 
2002ji & NGC3655 & 1538 & Ibc & ...  &\citet{2002IAUC.8028....3R} \\ 
2002jz & UGC 2984 & 1527 & Ic & ...  &\citet{2002IAUC.8037....2H}, somewhat uncertain \\ 
2003B & NGC1097 &1066 & II-P & ...  & M. Hamuy (priv. communication) \\ 
2003J & NGC4157& 1011 & II-P & ...  & \citet{2003IAUC.8048....2A}, amateur LC\\ 
2003Z &NGC2742 & 1518 & II-P & ...  & \citet{2007AaA...475..973U}, low luminosity II-P\\ 
2003bg & MCG-05-10-15 &1151 & Ic & ...  & \citet{2006ApJ...651.1005S}, evidence of H during evolution\\ 
2003bk & NGC4316 & 1325 & II & ...  &\citet{2003IAUC.8086....2P}, subtype unknown \\
2003ed & NGC5303 & 1631 & IIb & ...  &\citet{2003IAUC.8144....2L} \\ 
2003gd & NGC628 & 686 & II-P& in & \citet{2005MNRAS.359..906H} \\ 
2003hn & NGC1448 & 911 & II-P &out & M. Hamuy (priv. communication) \\ 
2003ie & NGC4051 & 917 & II-P& ...  & \citet{2008AaA...488..383H} \\ 
2003jg & NGC2997 & 914 & Ibc & in &\citet{2003IAUC.8241....2H} \\ 
2004A & NGC6207 & 1090 & II-P & in &\citet{Hen06} \\ 
2004C & NGC3683 & 1944 & Ic & ...  &\citet{2004IAUC.8269....2M} \\ 
2004G & NGC5668 & 1673 & II & out &\citet{2004IAUC.8273....2E}, subtype unknown \\ 
2004am & NGC3034 & 487 & II-P & in & Mattila in prep. \\
2004ao & UGC10862 & 1812 & Ib & ...  &\citet{2004IAUC.8304....4M} \\ 
2004bm & NGC3437 & 1381 & Ic & ...  &\citet{2004IAUC.8339....2F} \\ 
2004cm & NGC5486 & 1648 & II-P & ...  &\citet{2004IAUC.8359....1C}, probable low-luminosity II-P\\ 
2004cz &ESO407-G09 & 1398 & II-P & ...  & \citet{2004IAUC.8374....2F}\\ 
2004dg & NGC5806 & 1439 & II-P & in &\citet{2004IAUC.8376....2E}\\ 
2004dj & NGC2403 & 370 & II-P & out &\citet{2006MNRAS.369.1780V}, \citet{2006AJ....131.2245Z} \\ 
2004dk &NGC6118 & 1645 & Ib & ...  & \citet{2004IAUC.8404....1F} \\ 
2004ep &IC2152 & 1683 & II & ...  & \citet{2004IAUC.8420....2F}, subtype unknown \\
2004et & NGC6946 & 318 & II-P &...  & \citet{2007MNRAS.381..280M}, \citet{2006MNRAS.372.1315S}\\ 
2004ez & NGC3430 & 1731 & II-P & ...  &\citet{2004IAUC.8420....2F}, amateur LC \\ 
2004fc & NGC701 & 1727 & II-P & ...  &\citet{2004IAUC.8432....2S} \\ 
2004gk & IC3311 & -53 & Ic & ...  &\citet{2004IAUC.8446....1Q} \\ 
2004gn & NGC4527 & 1780& Ic & out & 1990B-like, SNIFS spectrum \footnote{http://www.supernovae.net/sn2004/sn2004gn.jpg}\\ 
2004gq & NGC1832 & 1782 & Ibc & - &\citet{2005IAUC.8461....3M}, \citet{2007AIPC..924..297G}\\ 
2004gt &NGC4038 & 1559 & Ibc & in & \citet{2005ApJ...630L..29G},\citet{Mau05}\\ 
2005V & NGC2146 & 1157 & Ibc & in &\citet{2005IAUC.8474....3T}\\ 
2005ad & NGC941 & 1535 & II-P & ...  &M. Hamuy (priv. communication), probable II-P\\ 
2005ae & E209-G09 & 862 & IIb& in & \citet{2005IAUC.8486....3F}\\ 
2005af & NGC4945 & 376 & II-P &...  & . \citet{2006AaA...454..827P} \\ 
2005at & NGC6744 & 632 & Ic &out & \citet{2005IAUC.8496....2S} \\ 
2005ay & NGC3938 & 1017 & II-P &...  & \citet{2008ApJ...685L.117G}\\ 
2005cs & NGC5194 & 702 & II-P &in & \citet{2006MNRAS.370.1752P}, low-luminosity II-P \\ 
2005cz &NGC4589 & 2283 & Ib & in &\citet{2005IAUC.8579....2L} \footnote{Although $V_{vir}$ outside  limit, the TF/SBF distance from \citet{2000ApJ...530..625T} puts it  within our distance limit and so it is included.}\\ 
2005kl & NGC4369 &1248 & Ic & ...  & \citet{2005CBET..305....1T} \\ 
2006bc & NGC2397 &1064 & II-P & in & \citet{2006CBET..450....1P}, amateur LC\\ 
2006bp & NGC3953 & 1285 & II-P & ...  &\citet{2008ApJ...675..644D} \\ 
2006jc & UGC4904 & 1830 & Ic & ..  &\citet{2007Natur.447..829P}, \citet{2007ApJ...657L.105F}: narrow He\footnote{SNe 2002ao and 2006jc have been termed Ibn as they show  narrow He lines due to circumstellar He rich shells. We adopt Ic rather than Ibn as that more accurately reflects the nature of the event.} \\ 
2006my & NGC4651 & 912.1 &II-P & in & \citet{Li07} \\ 
2006ov & NGC4303 & 1616 & II-P & in &\citet{Li07} \\ 
2007C & NGC4981 & 1687 & Ib & ...  &\citet{2007CBET..800....2B} \\ 
2007Y & NGC1187 & 1214 & Ibc & ...  &\citet{2007CBET..862....1F}\\ 
2007aa & NGC4030 & 1476 & II-P & in &\citet{2007CBET..850....1F}, our own LC\\ 
2007av & NGC3279 & 1436 &II-P & ...  & \citet{2007CBET..903....1H}, amateur LC\\ 
2007gr &NGC1058 & 634 & Ic & in & \citet{2008ApJ...673L.155V}\\ 
2007it &NGC5530 & 1045 & II & ...  & \citet{2007CBET.1068....1C}, unknown subtype \\ 
2007oc & NGC7418A & 1938 & II-P & ...  &\citet{2007CBET.1120....1O}, 1999em like \\ 
2007od & UGC12846 & 1810 &II-P & ...  & \citet{2007CBET.1119....1B}, 1999em like\\ 
2008M &E121-G26 & 1981 & II-P & ...  & \citet{2008CBET.1227....1F} \\ 
2008ax& NGC4490 & 797 & IIb & in & \citet{2008MNRAS.389..955P} \\ 
2008bk &NGC7793 & 60 & II-P & out & \citet{Mat08} \\ 
2008bo & NGC6643 & 1778 &Ib & ...  & \citet{2008CBET.1325....1N} \\ 
2008fb & UGC2813 & 1622 &II & ...  & \citet{2008CBET.1479....2K} \\ 
2008gz & NGC3672 & 1822 &II-P & ... & \citet{2011MNRAS.414..167R} \\ 
2008ij & NGC6643 & 1778 &II & out & \citet{2008CBET.1628....2C} \\ 
2008in & NGC4303 & 1616 &II-P & out & \citet{2011ApJ...736...76R} low luminosity II-P\\ 
2009G & IC4444 & 1812 &IIP & in & \citet{2009CBET.1664....1S} \\ 
2009H & NGC1084 & 1298 & IIP faint & in & \citet{Ben09} \\ 
2009N & NGC4487 & 1026 & IIP faint & in& \citet{Cha09} \\ 
2009at & NGC5301 & 1749 & II & ...  &\citet{2009CBET.1722....1H} \\ 
2009bw & UGC2890 & 1404 & II & ...  &\citet{2009CBET.1746....1S} \\ 
2009dd & NGC4088 & 989 & II & in &\citet{2009CBET.1765....1E} \\ 
2009dq & IC2554 & 1108 & II & ...  &\citet{2009CBET.1789....1A} \\ 
2009em & NGC157 & 1600 & Ic & out &\citet{2009CBET.1807....1F} \\ 
2009gj & NGC134 & 1404 & IIb & out &\citet{2009CBET.1858....1F} \\ 
2009hd & NGC3627 & 788 & II & in &\citet{2011ApJ...742....6E} \\ 
2009ib & NGC1559 & 1005 & IIP & in &\citet{Str09} \\ 
2009js & NGC918 & 1536 & IIP faint & out &\citet{2009CBET.1969....2S} \\ 
2009kr & NGC1832 & 1780 & IIP/L & in &\citet{Fra10b}, \citet{2010ApJ...714L.254E} \\ 
2009ls & NGC3423 & 1033 & IIL & in &\citet{Pej09} \\ 
2009md & NGC3389 & 1356 & II & in & \citet{Fra11}\\ 
2009mg & ESO121-G26 & 1981 & IIb & out &\citet{2010CBET.2158....1S} \\ 
2009mk & ESO293-34 & 1305 & IIb & ... &\citet{2009CBET.2086....1C} \\ 
2010br & NGC4051 & 917 & Ibc & in &\citet{2010CBET.2245....2M} \\ 
2010gi & IC4660 & 1538 & IIb & ... &\citet{2010CBET.2384....1Y} \\ 
2011am & NGC4219 & 1809 & Ib & in &\citet{2011CBET.2667....1M} \\ 
2011aq & NGC1056 & 1620 & II & ...  &\citet{2011CBET.2671....1C} \\ 
2011dh & NGC5194 & 702 & IIb & in &\citet{Mau11}, \citet{Van11} \\ 
2011dq & NGC337 & 1572 & IIP faint & &\citet{2011CBET.2749....1M} \\
2011hp		& NGC4219     		& 1811		& Ic  		& in	& \citet{2011CBET.2899....1M} 	\\
2011hs		& IC5267     		& 1524		& IIb 		& out	& \citet{2011CBET.2902....1M} 	\\
2011ja		& NGC4945     		& 376		& IIP 		& out	& \citet{2011CBET.2946....1M} 	\\
2011jm		& NGC4809     		& 979		& Ic  		& ...		& \citet{2011CBET.2962....1H} 	\\
2012A 		& NGC3239     		& 820		& II  		& ...		& \citet{2012CBET.2974....3S}	\\
2012P 		& NGC5806     		& 1439		& IIb		& in	& \citet{2012CBET.2993....2B}	\\
2012au 		& NGC4790     		& 1344		& Ib  		& in	& \citet{2012CBET.3052....2S}	\\		
2012aw		& NGC3351     		& 826		& IIP & in	& \citet{2012ApJ...759L..13F}, \citet{2012ApJ...756..131V}
\\
\\
{\bf Type Ia} \\
1998aq  	& NGC3982     		& 1388	& Ia     	& in   	&  \citet{2003AJ....126.1489B}\\
1998bn  	& NGC4462     		& 1718  	& Ia     	& ...  		&  \citet{1998IAUC.6888....1P} \\
1998bu  	& NGC3368     		& 941   	& Ia     	& out  	&  \citet{2000MNRAS.319..223H} \\
1998dm  	& PGC005341  	& 1882  	& Ia     	& ...  		& \citet{1998IAUC.6997....2F}\\
1999by 	& NGC2841       	& 831   	& Ia     	& ...     	&  \citet{2004ApJ...613.1120G} \\
2000E  	& NGC6951       	& 1704  	& Ia     	& ...     	&  \citet{2003ApJ...595..779V} \\
2001dp 	& NGC3953       	& 1285  	& Ia     	& out  	&  \citet{2001IAUC.7683....2A}      \\
2001el 	& NGC1448       	& 911   	& Ia     	& out   	&  \citet{2003AJ....125..166K}    \\ 
2001fu 	& M-03-23-11    	& 1564    	& Ia     	& ...     	&  \citet{2001IAUC.7748....3M}       \\
2002bo 	& NGC3190       	& 1410  	& Ia     	& out  	& \citet{2004MNRAS.348..261B}        \\  
2002cv 	& NGC3190       	& 1410  	& Ia     	& out  	&  \citet{2008MNRAS.384..107E}     \\
2002fk 	& NGC1309       	& 1986  	& Ia     	& ...    	&  \citet{2002IAUC.7976....3A}      \\
2003cg 	& NGC3169       	& 1237  	& Ia     	& in	  	&  \citet{2006MNRAS.369.1880E}     \\
2003gs 	& NGC936       		& 1279  	& Ia     	& in	  	&  \citet{2003IAUC.8173....3G}\\
2003hv 	& NGC1201      		& 1490  	& Ia     	& in    	&  \citet{2007ApJ...661..995G}  \\
2003hx 	& NGC2076      	 	& 1974  	& Ia     	& ...    	&  \citet{2008MNRAS.389..706M} \\
2003if 	& NGC1302       	& 1506  	& Ia     	& ... 	 	&  \citet{2003IAUC.8206....2M}   \\
2004W  	& NGC4649       	& 1198  	& Ia     	& in   	& \citet{2004IAUC.8286....2M} \\
2004ab 	& NGC5054       	& 1708  & Ia     	& in 	 	&  \citet{2004IAUC.8293....2M}     \\
2004ea 	& M-03-11-19    	& 1792  & Ia     	& ... 	 	&  \citet{2004IAUC.8409....2G}     \\
2005cf 	& M-01-39-03    	& 1977  & Ia     	& ...    	&  \citet{2007MNRAS.376.1301P} \\
2005cn 	& NGC5061       	& 1944  & Ia     	& out  	&  \citet{2005IAUC.8553....2M} \\
2005df 	& NGC1559       	& 1005  & Ia     	& in	  	&  \citet{2007ApJ...661..995G}      \\
2005ke 	& NGC1371       	& 1272  & Ia     	& ...	  	&  \citet{2006ApJ...648L.119I}      \\
2006E  	& NGC5338      	& 900   	& Ia     	& ...   	&  \citet{2006ATel..690....1A}   \\ 
2006X  	& NGC4321       	& 1677  & Ia     	& in    	&  \citet{2008ApJ...675..626W}  \\
2006ce 	& NGC908       		& 1344  & Ia     	& ...   	&  \citet{2006CBET..541....1B}   \\
2006dd 	& NGC1316     	 	& 1556  & Ia     	& in    	&  \citet{2006CBET..557....1S} \\
2006mq 	& ESO494-G26   	& 779.4   	& Ia     	&  .. 		&  \citet{2006CBET..731....1P} \\
2006mr 	& NGC1316      	& 1556  & Ia     	& in  		&  \citet{2006CBET..729....1P}  \\
2007af 	& NGC5584    		& 1704  & Ia   	& in 		&  \citet{2007ApJ...671L..25S}   \\   
2007bm 	& NGC3672    		& 1824  & Ia   	& ... 		&  \citet{2007CBET..939....1N}   \\
2007gi 	& NGC4036    		& 1629     & Ia   		& in  		& \citet{2007CBET.1021....1H} \\
2007le 	& NGC7721    		& 1971  & Ia   	& ... 		&  \citet{2007CBET.1101....1F}   \\
2007on 	& NGC1404    		& 1717  & Ia   	& in  		&  \citet{2008Natur.451..802V}  \\
2007sa 	& NGC3499    		& 1761  & Ia   	& ... 		&  \citet{2007CBET.1163....1A}   \\
2007sr 	& NGC4038    		& 1559  & Ia   	& in  		&  \citet{2007CBET.1174....1U}  \\
2008cq	& PGC731537  		& 1191		& Ia		& ...			&  \citet{2008CBET.1398....1M}\\
2008fp	& ESO428-G14		& 1504		& Ia		& in		&  \citet{2008CBET.1509....1W} \\								
2008ge	& NGC1527 			& 910		& Ia		& in		&  \citet{2010AJ....140.1321F} 2002cx-like\\								
2008ha	& UGC12682 			& 1469		& Ia		& ...			&  \citet{2009Natur.459..674V}, \citet{2009AJ....138..376F} 2002cx-like \\
2010ae	& ESO162-17			& 818		& Ia		& ...			&  \citet{2010CBET.2191....1S},  2002cx-like \\	
2010el	& NGC1566			& 1221 		& Ia		& in		&  \citet{2010CBET.2337....1B}, 2002cx-like \\	
2010ey	& ESO548-76			& 1312		& Ia		& ...			&  \citet{2010CBET.2355....1M} \\				
2010fz	& NGC2967			& 1861		& Ia		& ...			&  \citet{2010CBET.2364....1P} \\			
2010ih	& NGC2325			& 1954		& Ia		& out		&  \citet{2010CBET.2480....2M} \\	
2011B	& NGC2655			& 1653		& Ia		& out		&  \citet{2011CBET.2625....2N} \\	
2011ae	& MCG-03-30-19		& 1747		& Ia		& ...			&  \citet{2011CBET.2658....1H}		\\
2011at	& MCG-02-24-027		& 1922		& Ia		& ...			&  \citet{2011CBET.2676....1C} \\					
2011by	& NGC3972			& 1086		& Ia		& in		&  \citet{2011CBET.2708....2Z} \\	
2011dm	& UGC11861			& 1756		& Ia		& ...			&  \citet{2011CBET.2745....3M}\\	
2011ek	& NGC918			& 1536		& Ia		& out		& \citet{2011CBET.2785....1N} \\
2011fe	& M101				& 502		& Ia		& in		& \citet{2011CBET.2792....1N}, \citet{Li11}\\	
2011iv	& NGC1404			& 1717		& Ia 		& in		& \citet{2012ApJ...753L...5F}\\				
2011iy	& NGC4984			& 1180		& Ia 		& ...			& \citet{2011CBET.2943....2C} \\
\\
\\
{\bf Unclassified} \\
1998cf 		& NGC3504     	& 1669  	& ...    	& in 		& \\ %
1999gs 		& NGC4725      	& 1206 		& ...   		& out 		& \\
2008eh  	& NC2997 		& 91		& ...			& out 		& \\ 	 
2008iz		& NGC3034	 	& 487		& CC		& in		& \\ 	 																						
\\
{\bf Uncertain or}\\ {\bf non-supernova} \\
1999bw	 		& NGC3198   & 850.9 	& LBV   		& out	& \citet{1999IAUC.7152....2F}\footnote{http://etacar.umn.edu/etainfo/related/}\\ 
2000ch  		& NGC3432   & 778.6   	& LBV   		& out 	& \citet{2004PASP..116..326W} \\ 
2001ac 			& NGC3504   & 1668.9 	& LBV   		& out 	& \citet{2001IAUC.7597....3M}\\
2002kg  		& NGC2403   & 370		& LBV   		& out	& \citet{2006MNRAS.369..390M}\\
2003gm  		& NGC5334   & 1433.8 	& LBV   		& in  	& \citet{2006MNRAS.369..390M} \\
2007sv  		& UGC5979   & 1377.0 	& LBV    		&  ...  & \citet{2007CBET.1184....1H}, our own data \\
2006fp  		& UGC12182  & 1807.5   	& LBV   		&  ..   & \citet{2006CBET..636....1B} \\
M85 OT2006-1    & M85      	& 957      	& SN or OT?  	& in  	& \citet{2007Natur.447..458K} \\
2008S           & NGC6946  	& 318      	& IIn or OT? 	& out 	& \citet{2009MNRAS.398.1041B}  \\
NGC300 OT2008-1 & NGC300   	& 100      	& SN or OT?  	& in  	& \citet{2010MNRAS.408..181P} \\
2009ip			& NGC7259 	& 1596		& LBV			& in	& \citet{2012arXiv1210.3568P} and references within 	\\ 	 
2010U			& NGC4214	& 479		& Nova			& 	in	& \citet{2010ApJ...718L..43H} 	\\ 
2010da			& NGC300	& -38		& LBV/08S-like	&	in	& \citet{2011ApJ...739L..51B} 	\\ 
2010dn			& NGC3184	& 755		& LBV/08S-like	&	 in	& \citet{2010CBET.2300....1V} 	\\ 
2011ht		& UGC 5460     		& 			& IIn 		&  		& 	 IMPOSTOR\\

\end{longtable}
\label{lastpage}

\end{document}